\documentstyle[aps,prb,epsf,multicol,eqsecnum,amssymb]{revtex}

\newcommand{\bleq}{\ifpreprintsty
                   \else
                   \end{multicols}\vspace*{-3.5ex}{\tiny 
                   \noindent\begin{tabular}[t]{c|}
                   \parbox{0.493\hsize}{~} \\ \hline \end{tabular}}
                   \fi}
\newcommand{\eleq}{\ifpreprintsty
                   \else
                   {\tiny\hspace*{\fill}\begin{tabular}[t]{|c}\hline
                    \parbox{0.49\hsize}{~} \\ 
                    \end{tabular}}\vspace*{-2.5ex}\begin{multicols}{2}
                    \fi}
\newcommand{\bcols}{\ifpreprintsty\else\begin{multicols}{2}\fi}
\newcommand{\ecols}{\ifpreprintsty\else\end{multicols}\fi}

\begin{document}
\draft
\title{
Crossover from the chiral to the standard universality classes
in the conductance of a quantum wire with random hopping only
}

\author{Christopher Mudry,$^{a}$\cite{Chris} P.\ W.\ Brouwer,$^{a}$\cite{Piet}
and Akira Furusaki$^b$} 
\address{
$^a$Lyman Laboratory of Physics, Harvard University, Cambridge MA 02138\\
$^b$Yukawa Institute for Theoretical Physics, Kyoto University,
Kyoto 606-8502, Japan}

\date{\today}

\maketitle


\begin{abstract}
The conductance of a quantum wire with off-diagonal disorder that 
preserves a sublattice symmetry (the random hopping problem with
chiral symmetry) is considered. 
Transport at the band center is anomalous relative to the 
standard problem of Anderson localization both in the diffusive and 
localized regimes. In the diffusive regime, 
there is no weak-localization correction
to the conductance and universal conductance fluctuations are twice
as large as in the standard cases. 
Exponential localization occurs only for an even number of transmission 
channels in which case the localization length does not depend on 
whether time-reversal and spin rotation symmetry are present or not.
For an odd number of channels the conductance decays algebraically.
Upon moving away from the band center transport characteristics undergo
a crossover to those of the standard universality classes of 
Anderson localization. This crossover is calculated in the diffusive regime. 
Numerical simulations agree qualitatively with the theory.
\end{abstract}

%
\bcols

\section{Introduction}

Since the introduction of the scaling approach to
the problem of Anderson localization,\cite{Edwards72,Lee85} it is known
that transport characteristics of a disordered metal are universal,
provided the disorder is sufficiently weak, the temperature
sufficiently low so that quantum coherence is maintained over large
distances, and the interaction between electrons can be neglected.  An
example is the phenomenon of weak
localization,\cite{Anderson79,Gorkov79} a small deviation from Ohm's
law for the conductance of a weakly disordered metal, which is
suppressed by the application of a time-reversal symmetry breaking
magnetic field. Though small, the weak-localization correction is
universal in the sense that it does not depend on the shape of the
sample, nor on any other microscopic or macroscopic property other than
its dimensionality and the presence or absence of time-reversal
symmetry and spin-rotation invariance. Another example is the
phenomenon of universal conductance
fluctuations:\cite{Altshuler85,LeeStone85} The sample-to-sample
fluctuations of the conductance of a disordered metal or semiconductor
are of order $e^2/h$ with a prefactor that only depends on
dimensionality and symmetry. Both the weak-localization correction and
the universal conductance fluctuations are precursors of the true
Anderson localization, where as a result of destructive interference of
multiple scattered quantum mechanical waves the dirty metal turns into
an insulator for sufficiently strong disorder, or, in one or two
dimensions, for a sufficiently large sample size.\cite{ImryReview}

The original paper by Anderson,\cite{Anderson58} and most of the effort
devoted to the problem of Anderson localization since then, considers
the case of a particle on a lattice with a random on-site potential
(diagonal disorder) and non-random hopping amplitudes. In that case,
one distinguishes three universality classes, corresponding to the
presence or absence of time-reversal and of spin-rotation symmetry.
These three classes, are called orthogonal, unitary, and symplectic,
respectively. Here, we will refer to these as the three
``standard'' universality classes.

The electronic localization problem was soon generalized to lattice
models with randomness in the hopping amplitudes (off-diagonal
disorder).\cite{HerbertJones71} (This type of randomness was previously
known from the description of phonons\cite{Dyson53,Lifschitz64} and
narrow-gap semiconductors.\cite{Keldysh63}) The localization problem
with off-diagonal disorder has received comparatively much
less attention, although it has been known since the work of
Dyson\cite{Dyson53} that random systems with off-diagonal disorder, but
without diagonal disorder, can behave in a way dramatically different 
from that of systems with diagonal disorder only, or with both types of
disorder.\cite{Dyson53,Theo76,Fleishman77,Eggarter78,Oppermann79,Wegner81,Stone81}
For instance, the average density of states (DoS) for a one-dimensional
chain with random nearest-neighbor hopping was found to be singular at
the center of the band,
$\varepsilon=0$.\cite{Dyson53,Theo76,Eggarter78} According to the
Thouless formula,\cite{Thouless72} such a singular DoS implies that at
$\varepsilon=0$ the conductance distribution be anomalous as
well.\cite{Stone81,Mathur97} Wegner and Gade in Ref.~\onlinecite{Gade93} 
(see also Refs.~\onlinecite{Oppermann79,Hikami93,Fukui99,Altland99,Fabrizio00})
found a two-dimensional
counterpart to the singular behavior of the average DoS within their
analysis of a non-linear-$\sigma$ model with a sublattice
symmetry.
Interest in the effect of off-diagonal disorder has
revived in the 90's on two fronts. Motivated by quenched approximations
to interacting theories such as the quantum Hall effect at half-filling
or gauge approaches to high $T_c$ superconductivity, the random flux
problem (a special case of off-diagonal disorder in which hopping
amplitudes have a random phase only) has been extensively studied,
although very little consensus on its localization properties has
emerged.\cite{Furusaki99} A second thrust of activity has been
motivated by the close resemblance between the anomalies at zero
energy induced by pure off-diagonal disorder in two dimensions 
and the nature of the plateau transitions in the integer quantum Hall effect
(IQHE):\cite{Huckestein95} Both models might share the property that
all eigenstates are localized except at one special
energy.\cite{Furusaki99}

The reason why the localization properties of the random hopping
problem can depart from those of the standard problem of Anderson
localization is the existence of an additional sublattice symmetry in
systems with off-diagonal but without diagonal
disorder:\cite{Oppermann79,Wegner81,Ziman82,Inui94} In that case, the
lattice can be divided into two sublattices, such that the Hamiltonian
changes sign under a transformation where the wavefunction changes sign
on one sublattice, but not on the other. As a result, the spectrum is
symmetric with respect to a reflection about $\varepsilon=0$ (i.e.,
eigenvalues appear in pairs $\pm \varepsilon$). The fact that the band
center $\varepsilon=0$ is a very special energy in the presence of the
sublattice symmetry explains why anomalies in the DoS and
the localization properties occur at precisely this value of the
energy. When the energy moves away from zero, the effects of the
sublattice symmetry on the spectrum and the wavefunctions decreases 
and a crossover to the standard behavior takes place.
The sublattice symmetry is broken by the presence of on-site disorder,
long-range hopping,\cite{Inui94} or (in some cases) by periodic
boundary conditions.\cite{Mudry99} Counterparts to this sublattice
symmetry in other disordered systems or in quenched approximation to
interacting problems are numerous. They occur in, e.g., the QCD
Hamiltonian,\cite{QCD,Verbaarschot93} random $XY$
spin chains,\cite{McCoyWu68} diffusion in random
environments,\cite{random diffusion} supersymmetric quantum
mechanics,\cite{supersymQM} non-Hermitean quantum mechanics,\cite{NHQM}
and two-dimensional disordered models in the continuum such as Dirac
fermions with random vector potentials.\cite{Ludwig94}
Following previous works in this field, which adopted the nomenclature
of QCD,\cite{Verbaarschot93} we will refer to the sublattice symmetry
as {\em chiral} symmetry and will restrict our attention to random hopping
problems with this symmetry.

One-dimensional disordered systems with chiral symmetry have been
well-studied with all kinds of approaches and in various contexts (for
references, see the previous paragraph), and despite a continuing
confusion about semantics, their localization properties can be
considered well-understood. For two-dimensional systems the situation
is different (see Refs.~\onlinecite{Furusaki99,Eilmes98} 
and references therein).
Reliable analytical and numerical results are notoriously
hard to obtain, and no consensus has been reached to date, not even on
some most elementary issues. In view of this controversy, it is
particularly instructive to study the natural intermediate between
one and two dimensions, the thick (or ``quasi-one-dimensional'') 
disordered wire.
On the one hand, it shares the existence of both a localized and a
diffusive regime of quantum transport with two dimensional disordered
systems, while on the other hand, it allows for a controllable analytic
treatment, just like the truly one-dimensional system. Moreover, 
quasi-one-dimensional systems appear as a logical intermediate step in the
finite-size scaling approach for numerical simulations in two and three
dimensions.\cite{MacKinnon83}

Localization properties at the band center of a quasi-one-dimensional
quantum wire with off-diagonal disorder were investigated in several
previous publications by the authors, together with Simons and
Altland.\cite{Brouwer98,Mudry99,Brouwer99NON} In those
works we derived a chiral counterpart to the so-called DMPK
equation,\cite{BeenakkerReview,Dorokhov82,MPK88} a Fokker-Planck
equation that governs the distribution of the transmission eigenvalues
of a quantum wire without chiral symmetry.
Solution of the chiral DMPK equation for lengths beyond the localization
length of the standard DMPK equation showed that there is no exponential
localization if the number $N$ of propagating channels is odd (including
the one-dimensional case), while the conductance decays exponentially
with length if $N$ is even.
This parity effect is strikingly similar to the sensitivity of the
low-energy sector of a single antiferromagnetic spin-$N/2$ chain to the
parity of $N$,\cite{Haldabe83} on the one hand, or to the sensitivity
of the low-energy sector of $N$ coupled antiferromagnetic spin-1/2
chains to the parity of $N$,\cite{Dagotto96} on the other hand.
In the special case of the chiral Fokker-Planck equation without time
reversal invariance (random phase quantum wire), it was possible to
calculate exactly the crossover from the diffusive to the localized
regime for all moments of the conductance and to verify the validity of
the assumption of universality against a numerical simulation of the
random flux problem.\cite{Mudry99} 
The numerical simulations also confirmed that sufficiently far away from
the center of the band, transport is governed by the standard
universality classes.

A limitation of the approach relying on the Fokker-Planck equations for
the transmission eigenvalues is that it cannot describe how the
conductance distribution crosses over from the chiral to the standard
universality class as $\varepsilon$ is tuned away from zero. In the 
renormalization group language, 
each Fokker-Planck equation describes a fixed point
corresponding to a case of pure symmetry and the fixed points by
themselves cannot be used to infer how the scaling flows take place between
them. One possibility to obtain information about the crossover energy
and length scales below (above) which the physics is that of the chiral
(standard) universality classes, is to study the DoS of a chiral
quantum wire.\cite{Brouwer99DOS} However, unlike in the case of a
one-dimensional wire, where the Thouless formula connects conductance
and DoS, for a quasi-one-dimensional wire it is not
possible to infer transport properties from the DoS.
In this paper, we use an alternative approach, developed by one of us
for the study of transmission through a random waveguide with 
absorption.\cite{Brouwer98WAV} Focusing on
weak-localization corrections and universal conductance fluctuations,
we compute how, in the diffusive regime, the conductance distribution
of a quantum wire with random hopping crosses over from the chiral to
the standard universality classes as the energy is tuned away from zero.
We are not able to compute the crossover in the localized regime.
Instead, for the localized regime, we consider the conductance
distributions in the pure symmetry classes and compare to numerical
simulations to establish the crossover scale and to verify the validity
of our predictions.
 
The paper is organized as follows.  In Sec. \ref{sec:Microscopic model
and scattering matrix}, we define our microscopic model and derive the
symmetries of the scattering matrix in the presence of the chiral
symmetry. We then explain the scaling approach in Sec. \ref{sec:RG
approach}. The localized regime is studied in Sec. \ref{sec:localized
regime}. Our main results are presented in Sec. \ref{sec:Diffusive
regime}, where we consider the crossover from the chiral to the
standard universality classes in the diffusive regime. In Sec.\
\ref{sec:Numerical simulations} we compare our theoretical predictions
to a numerical simulation of a random hopping model on a square lattice.
We conclude in Sec.\ \ref{sec:Conclusions}.

\section{Microscopic model and scattering matrix}
\label{sec:Microscopic model and scattering matrix}

A convenient microscopic model that describes
a single particle hopping randomly between two sublattices is defined.
The symmetries obeyed by the scattering matrix associated to this
Hamiltonian are derived. Assuming weak disorder, the microscopic model
is approximated by a model defined in the continuum, for which the
scattering matrix is explicitly constructed.

\subsection{Microscopic lattice model with chiral symmetry}

In a general form, the Schr\"odinger equation for an
$N$-chain system with random hopping between two sublattices and 
without on-site randomness reads
\begin{equation}
-\varepsilon\Psi^{\vphantom{\dag}}(m)=
T^{\vphantom{\dag}}_{m  }\Psi^{\vphantom{\dag}}(m+1)+
T^{         {\dag}}_{m-1}\Psi^{\vphantom{\dag}}(m-1).
\label{eq:Schr eq}
\end{equation}
For a spinless particle,
$\Psi(m)$ is the $N$-component wavefunction where the index $m$ labels
the position along the chain. 
For a particle with spin-1/2,
$\Psi(m)$ is the $N$-component wavefunction made of spinors.
In that case, the $N\times N$ hopping matrix $T_m$ consists of
quaternions.\cite{Mehta}
The system, and the allowed hopping 
matrix elements are depicted in Fig.\ 1(a). Note that
the case of a square lattice
with nearest-neighbor random hopping is included in the general formula
(\ref{eq:Schr eq}), see Fig.\ 1(b).

In order to model transport, we consider a disordered region of finite
length $L = M a$, $a$ being the lattice constant, and attach ideal
leads with hopping matrix $T_m = {\openone_{N}}$ on both ends,
see Fig.\ 1(a).
Following Refs.\ \onlinecite{Brouwer98,Brouwer99NON}, we draw the
hopping matrices $T_m$ with $m$ inside the disordered region from a
distribution centered around the $N \times N$ unit matrix,
\begin{equation}
  T_m = \exp(\delta T_m).
\end{equation}
We distinguish three symmetry classes depending on the presence or
absence of time-reversal and spin-rotation symmetry. For a spinless
particle (or for a spin-1/2 particle in the presence of spin-rotation
symmetry), the hopping matrix $\delta T_m$ is real (complex) if
time-reversal symmetry is present (absent). These two cases are
commonly referred to as the orthogonal and unitary symmetry class and
are labeled by the symmetry index $\beta=1$ and $2$, respectively.
The case of broken spin-rotation symmetry with time-reversal
symmetry is denoted $\beta=4$ and is referred to as the symplectic class.
When $\beta=4$, the elements of the
$N\times N$ matrix $\delta T_m$ are real quaternions.\cite{Mehta}
The situation when both time-reversal symmetry and spin-rotation
symmetry are broken reduces to the unitary class ($\beta=2$) 
and will not be considered separately in this paper.
We further assume that $\delta T_m$ has a Gaussian
distribution, with zero mean and with variance given by
\begin{eqnarray}
\langle
 ({\delta T_{m }})^{\vphantom{\dag}}_{k l }
[({\delta T_{m'}})^{\vphantom{\dag}}_{k'l'}]^{\dag}
\rangle&=&
{2 \beta a  \over \gamma \ell}
\delta^{\ }_{mm'}
\nonumber \\ && \mbox{} \times
\left(
\delta^{\ }_{kk'}
\delta^{\ }_{ll'} 
-{1-\eta\over N}
\delta^{\ }_{kl}
\delta^{\ }_{k'l'}
\right),
\nonumber\\
\langle
 ({\delta T_{m }})^{\vphantom{\dag}}_{k l }
 ({\delta T_{m'}})^{\vphantom{\dag}}_{k'l'}
\rangle&=&
{2-\beta\over\beta}
\langle
 ({\delta T_{m }})^{\vphantom{\dag}}_{k l }
 [({\delta T_{m'}})^{\vphantom{\dag}}_{k'l'}]^{\dag}
\rangle, \nonumber \\
\label{eq: variance hopping}
\end{eqnarray}
where 
\begin{equation}
  \gamma = \beta N + 2 - \beta - {2(1-\eta)\over N}, 
\label{eq:gamma}
\end{equation}
and $\ell$ is the mean free path. (Why $\ell$ can be identified as the
mean free path is explained below.) 
Here, the symbol $\dag$ denotes the operation of complex conjugation
for $\beta=1,2$, whereas it denotes the operation of Hermitean conjugation
for quaternions for $\beta=4$.\cite{Mehta}
We assume weak disorder, $\ell \gg a$. 
The parameter $\eta$ governs the relative randomness of the
determinant of $T_m$. (See Ref.\ \onlinecite{Brouwer99NON} for the
reason for its introduction.)

We have chosen the statistical distribution (\ref{eq: variance
hopping}) for technical convenience; it allows for an exact solution of
the transport problem. As a justification for this choice, we recall
that the transport properties do not depend on details of the
microscopic model as long as disorder is weak, $\ell \gg a$, and the
length $L$ of the system is much larger than the mean free path. All
properties of the microscopic model are summarized in the two
parameters $\ell$ and $\eta$. [The proper value of the parameter $\eta$
depends on the details of the microscopic model under consideration.
For instance, for the random flux model\cite{Furusaki99} (which is a
special case of a random hopping model), $\eta=0$, while $\eta > 0$ in
generic random hopping models.\cite{Brouwer99NON}] To emphasize this
universality, we compare our final results to numerical simulations for
nearest-neighbor random hopping on a square lattice,
cf.\ Fig.\ 1(b).

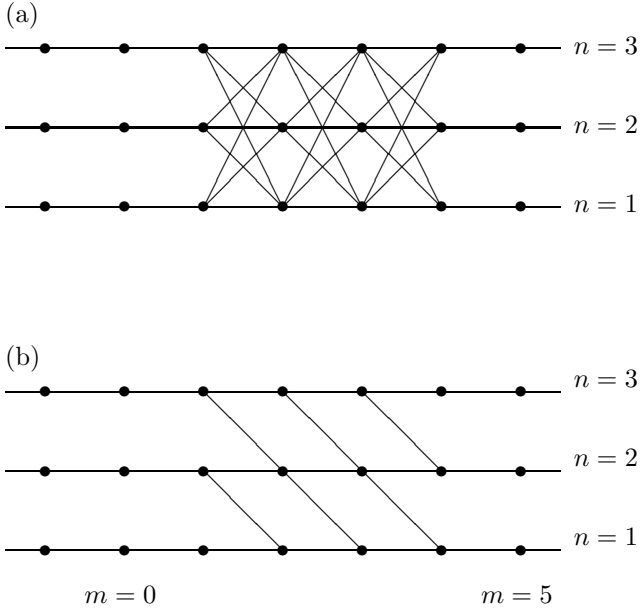
\begin{figure}
\begin{center}
\begin{picture}(240,250)(0, 0)
\put(  0,230){$({\rm a})$}
\put(  0,220){\line(1, 0){210}}\put(215,218){$n=3$}
\multiput(12.5,217.5)(30, 0){7}{$\bullet$}
\multiput(75,220)(30, 0){3}{\line(1,-1){30}}
\multiput(75,220)(30, 0){3}{\line(1,-2){30}}
\put(  0,190){\line(1, 0){210}}\put(215,188){$n=2$}
\multiput(12.5,187.5)(30, 0){7}{$\bullet$}
\multiput(75,190)(30, 0){3}{\line(1,-1){30}}
\multiput(75,190)(30, 0){3}{\line(1, 1){30}}
\put(  0,160){\line(1, 0){210}}\put(215,158){$n=1$}
\multiput(12.5,157.5)(30, 0){7}{$\bullet$}
\multiput(75,160)(30, 0){3}{\line(1, 1){30}}
\multiput(75,160)(30, 0){3}{\line(1, 2){30}}
\put(  0,100){$({\rm b})$}
\put(  0, 90){\line(1, 0){210}}\put(215, 92){$n=3$}
\multiput(12.5, 87.5)(30, 0){7}{$\bullet$}
\multiput(75  , 90  )(30, 0){3}{\line(1,-1){30}}
\put(  0, 60){\line(1, 0){210}}\put(215, 62){$n=2$}
\multiput(12.5, 57.5)(30, 0){7}{$\bullet$}
\multiput(75, 60)(30, 0){3}{\line(1,-1){30}}
\put(  0, 30){\line(1, 0){210}}\put(215, 32){$n=1$}
\multiput(12.5, 27.5)(30, 0){7}{$\bullet$}
%
%
%
\put(30, 10){$m=0$}\put(180, 10){$m=5$}
\end{picture}
\hfill\break
\end{center}
\caption{
(a) Random hopping model as described by Eq.\ (\protect\ref{eq:Schr eq}),
for $N = 3$. A disordered section of the wire (of length $M=4$) is
attached to ideal leads. Different chains are only coupled in the 
disordered region; 
there is no coupling between the chains in the perfect leads.
(b) The nearest-neighbor random hopping model on a
``square'' lattice is a special case of the model considered
under (a).
}
\label{fig:toy models}
\end{figure}

In the leads on the left (L) and right (R), the Schr\" odinger equation
(\ref{eq:Schr eq}) at energy $\varepsilon$ is solved by a sum of plane
waves moving towards the disordered region (denoted by a subscript
$\rm i$) and away from the sample (denoted by a subscript $\rm o$)
(see Fig.\ 2),
\begin{eqnarray}
  \Psi^{{\rm L}}_{\varepsilon}(m) &=& 
  \psi^{{\rm iL}}_{\varepsilon} e^{{i} k m a} +
  \psi^{{\rm oL}}_{\varepsilon} e^{-{i} k m a}, \nonumber \\
  \Psi^{{\rm R}}_{\varepsilon}(m) &=&
  \psi^{{\rm iR}}_{\varepsilon} e^{-{i} k m a} +
  \psi^{{\rm oR}}_{\varepsilon} e^{{i} k m a}.
\end{eqnarray}
Here $0 \le k \le \pi/a$, $\varepsilon = - 2 \cos k a$, and
$\psi^{\rm iL}_\varepsilon$ and 
$\psi^{\rm iR}_\varepsilon$ ($\psi^{\rm oL}_\varepsilon$ and
$\psi^{\rm oR}_\varepsilon$) are $N$-components vectors containing the
amplitudes of the incoming (outgoing) plane waves in the left and right
leads, respectively. The amplitudes of the ingoing and outgoing waves
are connected through the Schr\"odinger equation (\ref{eq:Schr
eq}) in the disordered region. This relation is formulated in terms of
the $2N \times 2N$ scattering matrix $S_\varepsilon$,
\begin{equation}
\pmatrix{
\psi^{\rm oL}_\varepsilon\cr
\psi^{\rm oR}_\varepsilon\cr
}
=
S^{\ }_\varepsilon\,
\pmatrix{
\psi^{\rm iL}_\varepsilon\cr
\psi^{\rm iR}_\varepsilon\cr}.
\label{eq:def S}
\end{equation}
Current conservation implies
\begin{eqnarray}
S^{\dag} S^{\vphantom{\dag}}=\openone_{2 N}.
\label{eq: flux cons S}
\end{eqnarray} 
(Here and below we suppress the index $\varepsilon$ if only scattering
matrices at the same energy are involved.)
For the cases $\beta=1,4$, i.e., if time-reversal symmetry is present,
the complex conjugate of any eigenfunction is itself an eigenfunction
with the same energy.
(For $\beta=4$, complex conjugation is meant in the
quaternion sense.\cite{Mehta}) 
Since outgoing and incoming plane waves are interchanged 
under complex conjugation, we infer that time-reversal invariance is
represented by the additional constraint
\begin{equation}
S^* S = \openone_{2 N}.
\label{eq: time rev on S}
\end{equation}

\begin{figure}
\begin{center}
\begin{picture}(240,100)(0,40)
\thicklines
\put(  0, 90){\line(1,0){240}}
\put(  0, 50){\line(1,0){240}}
\put(  0, 77){$  \psi^{\rm iL}_{\varepsilon n}  \longrightarrow$}
\put(200, 77){$\  \longrightarrow \psi^{\rm oR}_{\varepsilon n'}$}
\put(  0, 58){$  \psi^{\rm oL}_{\varepsilon n}  \longleftarrow $}
\put(200, 58){$\  \longleftarrow  \psi^{\rm iR}_{\varepsilon n'}$}
\put( 50, 50){\line(0,1){40}}
\put(190, 50){\line(0,1){40}}
\thinlines
\put(120, 40){\vector(-1,0){70}}
\put(120, 40){\vector( 1,0){70}}
\put(117, 30){$L$}
\put(50,85){\line(1,-1){35}}
\put(50,80){\line(1,-1){30}}
\put(50,75){\line(1,-1){25}}
\put(50,70){\line(1,-1){20}}
\put(50,65){\line(1,-1){15}}
\put(50,60){\line(1,-1){10}}
\put(50,55){\line(1,-1){ 5}}
\multiput(50,90)(5,0){21}{\line(1,-1){40}}
\put(190,85){\line(-1,1){ 5}}
\put(190,80){\line(-1,1){10}}
\put(190,75){\line(-1,1){15}}
\put(190,70){\line(-1,1){20}}
\put(190,65){\line(-1,1){25}}
\put(190,60){\line(-1,1){30}}
\put(190,55){\line(-1,1){35}}
\end{picture}
\hfill\break
\end{center}
\caption{
Quantum wire with a disordered region of length $L=Ma$.
Incoming plane waves are 
$\psi^{\rm iL}_{\varepsilon n }$ and 
$\psi^{\rm iR}_{\varepsilon n'}$.
Outgoing plane waves are 
$\psi^{\rm oL}_{\varepsilon n }$ and 
$\psi^{\rm oR}_{\varepsilon n'}$.
There are $N$ channels, i.e., $n,n'=1,\cdots,N$.
In a quasi-one-dimensional geometry, $L=Ma\gg Na$.
}
\label{fig:wire}
\end{figure}
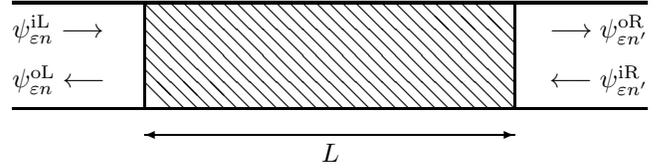

The Schr\"odinger equation (\ref{eq:Schr eq}) has an additional
symmetry: the Hamiltonian changes sign under the transformation
$\Psi(m) \rightarrow(-)^m \Psi(m)$. Correspondingly, for any
realization of the disorder, the spectrum of energy eigenvalues is
symmetric about the band center $\varepsilon=0$. This symmetry, which
originates from the fact that the disorder preserves the bipartite
structure of the lattice, is referred to as chiral symmetry.  The
chiral symmetry is a special attribute of random hopping between
different sublattices; it is broken by e.g.\ on-site randomness or
next-nearest-neighbor hopping.
It is the chiral symmetry that is responsible for the anomalous
transport properties at the special energy $\varepsilon=0$ of
a quantum wire with random hopping.\cite{Stone81,Mathur97,Brouwer98} 
To find the effect of the chiral symmetry on the scattering matrix, 
we note that the transformation $\Psi(m) \rightarrow(-)^m \Psi(m)$
interchanges incoming waves at energy $\varepsilon$ into outgoing
waves at energy $-\varepsilon$, and vice versa. Applied to Eq.\
(\ref{eq:def S}), this gives 
\begin{eqnarray}
\pmatrix{
\psi^{\rm iL}_{-\varepsilon}\cr
\psi^{\rm iR}_{-\varepsilon}\cr
}
&&=
S^{\ }_\varepsilon\,
\pmatrix{
\psi^{\rm oL}_{-\varepsilon}\cr
\psi^{\rm oR}_{-\varepsilon}\cr
}
=
\left(S^{\ }_{-\varepsilon}\right)^{-1}\,
\pmatrix{
\psi^{\rm oL}_{-\varepsilon}\cr
\psi^{\rm oR}_{-\varepsilon}\cr
}.
\end{eqnarray}
(The second equality follows from Eq.\ (\ref{eq:def S}) at energy
$-\varepsilon$.)
Taken together with flux conservation (\ref{eq: flux cons S}), we thus
find that the presence of the chiral symmetry results in the
constraint
\begin{equation}
S^{\vphantom{\dag}}_{\varepsilon}
=\left(S^{\ }_{-\varepsilon}\right)^{\dag}
\label{eq: chiral sym on S}
\end{equation}
for the scattering matrix $S$. Unlike the constraints of flux
conservation and time-reversal symmetry,
Eq.\ (\ref{eq: chiral sym on S}) involves 
scattering matrices at different energies. The exception is the band
center $\varepsilon=0$, where we find that $S$ is Hermitian,
\begin{equation}
  S^{\vphantom{\dag}}_0 = S_0^{\dagger}. 
\label{eq:Schiral}
\end{equation}

The scattering matrix is decomposed into four $N \times N$ subblocks
$r$, $r'$ and $t$, $t'$, the reflection and transmission matrices,
\begin{equation}
  S =
\pmatrix{
r&t'\cr
t&r'\cr
}.
\label{eq:def rt}
\end{equation}
The transmission and reflection matrices determine the transport
properties of the quantum wire. They are related to the conductance 
of the wire through the Landauer formula,
\begin{equation}
G= 
{2 e^2\over h}\, \mbox{tr}\, t^{\dagger} t\equiv
{2 e^2\over h}\, g, \label{eq:Landauer}
\end{equation}
and to the shot noise power\cite{Buettiker90}
\begin{equation}
P = 
{4 e^3 V \over h}\, \mbox{tr}\left[t^{\dagger} 
t({1} - t^{\dagger} t)\right]\equiv
{4 e^3 V \over h}\, p, \label{eq:ShotNoise}
\end{equation}
$V$ being the applied voltage. 
(See Ref.\ \onlinecite{BeenakkerReview}
for more applications to quantum transport.)
A further decomposition of $S$ follows from the polar decomposition of
the matrices $r,r'$ and
$t,t'$,
\begin{equation}
S = \pmatrix{
{\cal V}'&0\cr
0&{\cal U}\cr
}
\pmatrix{
 \tanh X        &(\cosh X)^{-1} \cr
 (\cosh X)^{-1} &- \tanh X      \cr
}
\pmatrix{
{\cal V}&0\cr
0&{\cal U}'\cr
},
\label{eq:polar decom of S}
\end{equation}
where 
${\cal U}$, ${\cal U}'$, ${\cal V}$, and ${\cal V}'$ 
are $N\times N$ unitary matrices and $X$ is an $N \times N$ diagonal
matrix with real numbers $x_j$ ($j=1,\cdots,N$) on the diagonal.
In the presence of time-reversal symmetry, one has
\begin{eqnarray}
&&
{\cal U}^*{\cal U}'={\cal V}^*{\cal V}'=\openone^{\ }_{N}.
\end{eqnarray}
Chiral symmetry implies a relationship between the unitary matrices
${\cal U}$, ${\cal U}'$, ${\cal V}$, and ${\cal V}'$ at opposite
energies,
\begin{eqnarray}
&&
{\cal U}^{\ }_{\varepsilon}={\cal U}'^{\dagger}_{-\varepsilon},
\quad
{\cal V}^{\ }_{\varepsilon}={\cal V}'^{\dagger}_{-\varepsilon},
\quad
X^{\ }_{\varepsilon}=X^{\ }_{-\varepsilon}.
\end{eqnarray}
In terms of the eigenvalues $x_j$, the equations (\ref{eq:Landauer})
and (\ref{eq:ShotNoise}) for the conductance and the shot noise power
read
\begin{equation}
  g = \sum_{j=1}^{N} {1 \over \cosh^2 x_j},
\qquad
  p = \sum_{j=1}^{N} {\tanh^2 x_j \over \cosh^2 x_j}.
  \label{eq:LandauerX}
\end{equation}

\subsection{Continuum model with chiral symmetry}

For weak disorder (mean free path $\ell$ much larger than the lattice
spacing $a$), we may replace the lattice model (\ref{eq:Schr eq}) by
a continuum model.
We linearize the spectrum of the kinetic energy of Schr\"odinger
equation (\ref{eq:Schr eq}) in the close vicinity of the band center
$\varepsilon=0$.
Choosing a representation with left and right
movers, we arrive at the continuum Schr\"odinger equation
\begin{equation}
-\varepsilon\psi (y) =
\left[
{i}\sigma_3\otimes \openone_{N}\partial_y+
\sigma_3\otimes v(y)+
\sigma_2\otimes w(y)
\right]\psi(y).
\label{eq:Schr eq cont}
\end{equation}
Here $\psi$ is a $2N$ component vector (elements of $\psi$ occur in
pairs that correspond to left and right movers), $v$ and $w$ are
$N \times N$ Hermitean matrices, and the $\sigma_\mu$ $(\mu = 1,2,3)$
are the Pauli matrices.
In the presence of time-reversal symmetry $w$ ($v$) is (anti)symmetric. 
The continuum limit has been taken along the chains only; discreteness
is maintained in the transverse direction through the $N$ components of
$\psi$. The Fermi velocity has been set to one.
The randomness in the hopping amplitudes has been translated to the
matrices $v$ and $w$, by means of the identifications
\begin{eqnarray}
{i}
\left(
\delta T^{\hphantom{\dag} }_m
-
\delta T^{\dag}_{m+1}
\right)+\hbox{ h.c. } 
&\rightarrow& v(y),
\nonumber\\
-
\left(
\delta T^{\hphantom{\dag} }_m
-
\delta T^{\dag}_{m+1}
\right)+\hbox{ h.c. } 
&\rightarrow& w(y).
\label{eq: identifications v w}
\end{eqnarray} 
With the choice (\ref{eq: variance hopping}), the disorder in $v$ 
is statistically independent from the disorder in $w$. 
Both $v$ and $w$
are Gaussian distributed with zero mean and with variance
\begin{mathletters}
\label{eq: stat of v and w}
\begin{eqnarray}
\left\langle
v^{\ }_{ij}(y) v^{\ }_{kl}(y')^{\dag}
\right\rangle &&=
{\beta\, \delta(y-y')\over \gamma \ell} 
\bigg[
\delta^{\ }_{ik}\delta^{\ }_{jl}
-
{2-\beta\over\beta}
\delta^{\ }_{il}\delta^{\ }_{jk}
\nonumber\\
&&
-{2(\beta-1)(1-\eta)\over\beta N}
\delta^{\ }_{ij}\delta^{\ }_{kl}
\bigg],
\\
\left\langle
w^{\ }_{ij}(y) w^{\ }_{kl}(y')^{\dag}
\right\rangle &&=
{\beta\, \delta(y-y') \over \gamma \ell}
\bigg[
\delta^{\ }_{ik}\delta^{\ }_{jl}
+
{2-\beta\over\beta}
\delta^{\ }_{il}\delta^{\ }_{jk} \nonumber \\ &&
-
{2(1-\eta)\over\beta N}
\delta^{\ }_{ij}\delta^{\ }_{kl}
\bigg]. 
\end{eqnarray}
\end{mathletters}

The symmetries (flux conservation, time-reversal, and chiral symmetry)
of the scattering matrix in the continuum model are the same as for 
the lattice model. (Note that in the continuum model, the chiral 
transformation is represented by $\psi\rightarrow\sigma_1\psi$. The
chiral symmetry then follows from the fact that $\sigma_1$ anticommutes
with the Hamiltonian.)

\section{Scaling approach}
\label{sec:RG approach}

The idea\cite{Anderson80} behind the scaling
approach to the theory of localization in a quantum wire is to
calculate how the scattering matrix $S$ of the quantum wire changes if
a thin slice is added to the disordered region [see Fig.\ 3(a)].
Here we are mostly interested in the eigenvalues of the matrix product
$t^{\dagger} t = {1} - r^{\dagger} r$, 
i.e., in the parameters $x_j$ of
the decomposition (\ref{eq:polar decom of S}). Hence, it is sufficient
to consider the reflection matrix $r$, and calculate how it is changed
upon the addition of a thin slice.
This change follows from the composition law
\begin{equation}
r^{\ } =
r^{\ }_1+t'_{1}\left({1}-r^{\ }_2r'_1\right)^{-1} r^{\ }_2t^{\ }_1,
\label{eq:r12}
\end{equation}
that gives the reflection matrix of two scatterers $1$ and $2$ in
series, in terms of the reflection matrix of the right scatterer
(2) and all reflection and transmission matrices of the left scatterer
(1), see Fig.\ 3(b). 

\begin{figure}
\begin{center}
\begin{picture}(240,200)(0,-180) 
\put(0,10){(a)}
\thicklines
\put(  0,-10){\line(1,0){240}}
\put(  0,-50){\line(1,0){240}}
%
%
\put( 50,-50){\line(0,1){40}}
\put(190,-50){\line(0,1){40}}
%
%
\thinlines
\put(120,-60){\vector(-1,0){70}}
\put(120,-60){\vector( 1,0){70}}
\put(117,-70){$L$}
\thicklines
\put( 40,-50){\line(0,1){40}}    
\thinlines
\put( 20,  0){\vector( 1,0){20}} 
\put( 70,  0){\vector(-1,0){20}} 
\put( 40,  5){$\delta L$}        
\put(50,-15){\line(1,-1){35}}
\put(50,-20){\line(1,-1){30}}
\put(50,-25){\line(1,-1){25}}
\put(50,-30){\line(1,-1){20}}
\put(50,-35){\line(1,-1){15}}
\put(50,-40){\line(1,-1){10}}
\put(50,-45){\line(1,-1){ 5}}
\multiput(50,-10)(5,0){21}{\line(1,-1){40}} 
\put(190,-15){\line(-1,1){ 5}}
\put(190,-20){\line(-1,1){10}}
\put(190,-25){\line(-1,1){15}}
\put(190,-30){\line(-1,1){20}}
\put(190,-35){\line(-1,1){25}}
\put(190,-40){\line(-1,1){30}}
\put(190,-45){\line(-1,1){35}}
%
 \multiput( 40,-50)(0,5){7}{\line(1,1){10}} 
 \multiput( 50,-50)(0,5){7}{\line(-1,1){10}} 
%
%
\put(  0,-100){(b)}
%
%
\thicklines
\put(  0,-120){\line(1,0){240}}
\put(  0,-160){\line(1,0){240}}
%
%
\multiput(30,-160)(40,0){2}{\line(0,1){40}} 
\multiput(170,-160)(40,0){2}{\line(0,1){40}} 
\thinlines
%
%
\multiput(70,-160)(-4,0){8}{\line(-1,1){10}} 
\multiput(30,-120)(4, 0){8}{\line(1,-1){10}} 
\multiput(30,-120)(0,-4){8}{\line(1,-1){10}} 
\multiput(70,-160)(0, 4){8}{\line(-1,1){10}} 
%
%
\multiput(210,-160)(-4,0){8}{\line(-1,1){10}} 
\multiput(170,-120)(4, 0){8}{\line(1,-1){10}} 
\multiput(170,-120)(0,-4){8}{\line(1,-1){10}} 
\multiput(210,-160)(0, 4){8}{\line(-1,1){10}} 
%
%
\put(  0,-133){$t'    _1\leftarrow    $}
\put(  0,-150){$r^{\ }_1\hookleftarrow$}
\put(218,-133){$\rightarrow     t^{\ }_2$}
\put(218,-150){$\hookrightarrow r'    _2$}
\put( 75,-133){$\rightarrow     t^{\ }_1$}
\put( 75,-150){$\hookrightarrow r'    _1$}
\put(140,-133){$t'    _2\leftarrow    $}
\put(140,-150){$r^{\ }_2\hookleftarrow$}
%
%
\put( 45,-142){$S_1$}
\put(185,-142){$S_2$}
%
%
\end{picture}
\hfill\break
\end{center}
\caption{
(a) A thin slice of length $\delta L$ with $a\ll\delta L\ll\ell\ll L$
is added to the left of the disordered region of length $L$. 
(b) Two disordered regions 1 and 2 with scattering matrices $S_1$ and
$S_2$, respectively, in a quantum wire.
}
\label{fig:composition law and scaling step}
\end{figure}
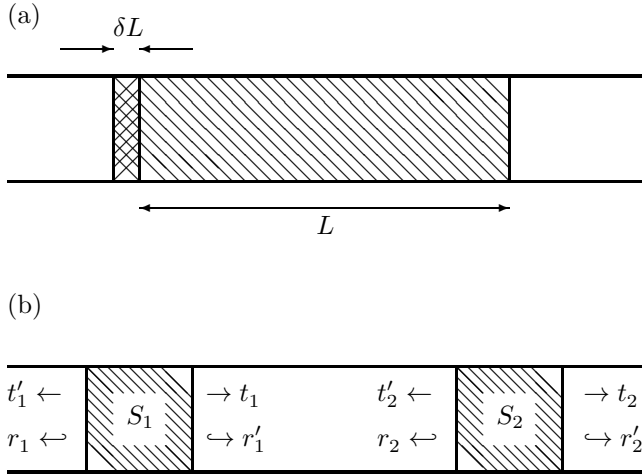

If applied to a quantum wire, the only input in this approach is the
statistical distribution of the transmission and reflection matrices
$t^{\ }_1$, $t_1'$, $r^{\ }_1$, and $r'_1$ of the thin slice. 
The width $\delta L$ 
of the slice is taken much smaller than the mean free path $\ell$,
so that the change of $r$ is small as well,
although $\delta L$ must remain large compared to the lattice spacing
$a$ for the continuum limit to be a good approximation. 
Then, the scattering matrix
$S_1$ of the thin slice can be calculated in the second-order Born
approximation from the Schr\"odinger Equation (\ref{eq:Schr eq cont}).
The result is
\begin{mathletters}
\label{eq: building block for S}
\begin{eqnarray}
r^{\ }_{1}&&=
-W+\case{{i}}{2}[V,W],
\\
t^{\ }_{1}&&=
{1}+
{i} V -
\case{1}{2}V^2-\case{1}{2}W^2 + {i} \varepsilon \delta L,
\label{eq: t slice}
\\
r'_{1}&&=
W+\case{{i}}{2}[V,W],
\\
t'_{1}&&=
{1} -
{i} V -
\case{1}{2}V^2-\case{1}{2}W^2 + {i} \varepsilon \delta L,
\label{t' slice}
\end{eqnarray}
\end{mathletters}
where
$$
V=\int_0^{\delta L}\!\!dy\, v(y),
\qquad
W=\int_0^{\delta L}\!\!dy\, w(y).
\label{eq: V and W}
$$
Here we neglected terms that are of order $(\delta L)^2$. [We also
ignored the $y$-ordering of the integrals in Eq.\ (\ref{eq:  building
block for S}) as it does not affect the statistical distribution of
$S_1$ in view of the delta-function correlation of the random
potentials $v$ and $w$.] Using Eq.\ (\ref{eq: stat of v and w}) for the
distribution of the random potentials $v$ and $w$, we find that the
matrices $V$ and $W$ are Gaussian distributed with zero average and
with variance proportional to the width $\delta L$ of the thin slice,
\begin{mathletters}
\label{eq: stat of slice}
\begin{eqnarray}
\left\langle
V^{\ }_{ij}(V^{\ }_{kl})^{\dag}
\right\rangle &&=
{\beta\delta L\over\gamma\ell}
\bigg[
\delta^{\ }_{ik}\delta^{\ }_{jl}
-
{2-\beta\over\beta}
\delta^{\ }_{il}\delta^{\ }_{jk}
\nonumber\\
&&
-{2(\beta-1)(1-\eta)\over\beta N}
\delta^{\ }_{ij}\delta^{\ }_{kl}
\bigg],
\\ 
\left\langle
W^{\ }_{ij}(W^{\ }_{kl})^{\dag}
\right\rangle &&=
{\beta\delta L\over\gamma\ell}
\Big[
\delta^{\ }_{ik}\delta^{\ }_{jl}
+
{{2-\beta\over\beta}}
\delta^{\ }_{il}\delta^{\ }_{jk}
\nonumber\\
&&-
{2\left(1-\eta\right)\over \beta N}\delta^{\ }_{ij}\delta^{\ }_{kl}
\Big].
\label{eq: var W} 
\end{eqnarray}
\end{mathletters}
Equations (\ref{eq:r12})--(\ref{eq: stat of slice}) define the scaling
approach. They are exact for the continuum model (\ref{eq:Schr eq
cont}) with the statistical distribution (\ref{eq: stat of v and w}) of
the random potentials, which in turn was derived from the random
hopping lattice model (\ref{eq:Schr eq},\ref{eq: variance hopping}) in
the limit of weak disorder. A different choice for the distribution of
the hopping matrices in Eq. (\ref{eq: variance hopping}) would have
led to different statistical properties of the scattering matrix for a
thin slice. However, as we will verify 
in Sec.\ \ref{sec:Numerical simulations} 
by numerical simulations, such differences are irrelevant in the sense
of the renormalization group,
i.e., they disappear for sufficiently long  wires (longer than the mean
free path $\ell$).

Note that the reflection probability 
$N^{-1} \mbox{tr}\, r_1^{\dagger} r_1^{\vphantom{\dagger}}$
of a thin slice has average
\begin{equation}
 N^{-1} \langle\mbox{tr}\, r_1^{\dagger} r^{\vphantom{\dagger}}_1 \rangle 
= \delta L/\ell,
\label{eq:Born}
\end{equation}
which justifies our choice that $\ell$ is the mean free path.

In terms of the matrices $V$ and $W$, upon addition of the thin slice,
the reflection matrix $r$ changes according to
\begin{mathletters}
\label{eq: r to r + delta r}
\begin{equation}
r\rightarrow r+\delta r,
\end{equation}
with
\begin{eqnarray}
\delta r&=&
 2{i}\varepsilon\delta Lr
-W+rWr-{i}(Vr-rV)+rWrWr
 \nonumber \\ && \mbox{}
-\case{1}{2}(W^2r + rW^2 + V^2 r  + r V^2)
+VrV.
\label{eq:delta r}
\end{eqnarray}
\end{mathletters}
We have not included terms of order $VW$ as
their contributions vanish upon disorder averaging.

Several observations can be made already on the level of the evolution
equation (\ref{eq: r to r + delta r}), in combination with the Gaussian
distribution (\ref{eq: stat of slice}) of the matrices $V$ and $W$. 
First, the distribution of $r$ is symmetric under a change of sign, $r
\to -r$. This implies that the average of any odd function of $r$
must be zero, for all values of the energy $\varepsilon$. 

Second, at the band center $\varepsilon = 0$, the chiral symmetry
implies that $r$ is Hermitian, cf.\ Eq.\ (\ref{eq:Schiral}). The
Hermiticity is broken by the first term in Eq.\ (\ref{eq:delta r}),
which is proportional to the energy. 

Third, the distribution of $r$ is invariant under transformations $r
\to U r U^{\dagger}$, where $U$ is an orthogonal (unitary) $N \times N$
matrix for $\beta=1$ ($2$). For zero energy, where $r$ is Hermitian,
this implies that the distribution of $r$ depends on its eigenvalues
$\tanh x_j$ only, cf.\ Eq.\ (\ref{eq:polar decom of S}). As was shown
in Refs.\ \onlinecite{Brouwer98,Brouwer99NON}, in this case, the scaling
flow can be represented in terms of a Fokker-Planck equation for the
distribution $P(x_1,\cdots,x_N;L)$,
\begin{eqnarray}
&&
{\partial P\over\partial L}=
{1\over \gamma \ell}\sum_{i,j=1}^N
{\partial\over\partial x_i}
\left(
\delta^{\ }_{ij}-{1-\eta\over N}
\right)
J{\partial\over\partial x_j} J^{-1}P,
\nonumber
\\
&&
J=\prod_{k<l}|\sinh(x_l-x_k)|^\beta.
\label{eq:FP ch}
\end{eqnarray}
Away from the center of the band, $r$ is no longer Hermitian, and its
distribution depends on both eigenvalues and eigenvectors.  However,
for $\varepsilon$ sufficiently far away from $0$ (this notion will be
made precise below), the chiral symmetry has no effect on the
scattering matrix, and $P(x_1,\cdots,x_N;L)$ obeys the Fokker-Planck
equation for the standard orthogonal, symplectic, or unitary symmetry 
classes, the so-called Dorokhov-Mello-Pereyra-Kumar (DMPK) equation,
\cite{Dorokhov82,MPK88}
\begin{eqnarray}
&&
{\partial P\over\partial L}=
{1\over2(\beta N + 2 - \beta)\ell}\sum_{j=1}^N
{\partial\over\partial x_j}
J{\partial\over\partial x_j} J^{-1}P,
\nonumber
\\
&&
J=
\prod_k|\sinh 2x_k|\prod_{k<l}|\sinh^2x_l-\sinh^2x_k|^\beta.
\label{eq:FP DMPK}
\end{eqnarray}
There is no parameter $\eta$ in the DMPK equation; the presence of the
parameter $\eta$ is special for the case of chiral symmetry at the band
center $\varepsilon=0$. In the language of the Fokker-Planck equation
(\ref{eq:FP ch}), $\eta$ controls the relative strength of the
diffusion of the center of mass $\bar x=(x_1+\cdots+x_N)/N$ compared to
that of the relative coordinates $x_j-\bar x$.

The most important difference between the Fokker-Planck equations
(\ref{eq:FP ch}) and (\ref{eq:FP DMPK}) are the symmetries of the
Jacobians $J$. In Eq.\ (\ref{eq:FP ch}), i.e., at the band center
$\varepsilon=0$, $J$ is invariant under a simultaneous translation $x_j
\to x_j + \delta x$ and under a simultaneous reflection
$x_j\rightarrow -x_j$ for all $j$. [The translation invariance decouples
the motion of the ``center of mass'' $\bar x=(x_1+\cdots+x_N)/N$ from
the relative coordinates $x_j-\bar x$, and hence calls for the presence
of the parameter $\eta$ in Eq.\ (\ref{eq:FP ch}).] 
In the standard DMPK equation (\ref{eq:FP DMPK}), i.e., 
for energies $\varepsilon$ far away from the
band center, $J$ is invariant under a reflection $x_j \to-x_j$ for each
$j$ separately; there is no longer translation invariance.  It is the
absence of this ``local'' reflection symmetry at $\varepsilon=0$ that
is responsible for anomalies in transport properties at
$\varepsilon=0$. In the remainder of this paper, we describe these in
more detail, focusing on the distribution of the conductance in
the localized regime $L \gg N \ell$ and on the quantum interference
corrections to the conductance in the diffusive regime $\ell \ll L \ll
N \ell$. For the localized regime, we use the Fokker-Planck equations
(\ref{eq:FP ch}) and (\ref{eq:FP DMPK}) to compare the transport
properties for $\varepsilon=0$ and $\varepsilon$ far away from $0$. (A
comparison for the case of broken time-reversal symmetry only has
already been given in Ref.\ \onlinecite{Mudry99}.) In the diffusive
regime we start from the evolution equation (\ref{eq: r to r + delta
r}) directly, in order to include the $\varepsilon$-dependence of the
transport properties. Knowledge of the crossover as a function of
$\varepsilon$ will allow us to specify what is meant by ``$\varepsilon$
sufficiently far away from $0$'', and hence when the standard DMPK
equation  (\ref{eq:FP DMPK}) replaces the special
Fokker-Planck equation (\ref{eq:FP ch}) in the random hopping problem.

\section{localized regime}
\label{sec:localized regime}

Differences between the conductance distribution at the band center
$\varepsilon=0$ and away from $\varepsilon=0$ are most pronounced in
the localized regime $L \gg N \ell$. Away from the band center, the
conductance decreases exponentially with length, as is the case in the
standard orthogonal, symplectic, and unitary classes. 
At the band center, however,
the exponential decrease of the conductance is only observed if the
number of channels is even, while for an odd number of channels the
conductance decreases only algebraically.\cite{Brouwer98} 

Exact calculations for the moments of the conductance in the standard
symmetry classes have been obtained for all
$\beta$,\cite{Zirnbauer92,BeenakkerRejaei,Mirlin94,Frahm95,BrouwerFrahm96}
while for the chiral symmetry classes governed by the Fokker-Planck
equation (\ref{eq:FP ch}) only exact results for $\beta=2$ and $\eta=1$
are known.\cite{Mudry99} While we do not know of a way to extend our
exact analysis of Ref.\ \onlinecite{Mudry99} to the cases of orthogonal
and symplectic symmetries, it is still possible to extract
the conductance distribution deep inside
the localized regime $L \gg N \ell$ using the approximation scheme of
Refs.\ \onlinecite{Dorokhov82,Dorokhov83,Pichard91}. This is done here.
We are thus able to compare the average and variance of the conductance,
and the average and variance of its logarithm at and away from the
band center $\varepsilon=0$ for the orthogonal, symplectic, and  unitary 
symmetry classes for all values of $\eta$.

Our starting point is the Fokker-Planck equation (\ref{eq:FP ch}),
which we rewrite in the form
\begin{equation}
  {\partial P\over\partial L} =
  {1\over\gamma \ell}\sum_{i,j=1}^N
  \left(\delta_{ij} - {1-\eta \over N} \right)
  {\partial\over\partial x_i}
  \left( {\partial P\over\partial x_j} +
  \beta P{\partial\Omega\over\partial x_j} \right)
\label{eq:FP bis}
\end{equation}
with the ``potential''
\begin{equation}
\Omega=
-\sum_{k=1}^N\sum_{l=k+1}^N \ln|\sinh(x_k-x_l)|.
\end{equation}
Equation\ (\ref{eq:FP bis}) has the interpretation that as $L$ increases,
fictitious particles with the coordinates $x_j$ perform a Brownian
motion subject to the repulsive two-body potential $\Omega$. Since
$\Omega$ has a hard core, we may assume that $x_1<x_2<\cdots<x_N$ for
all $L$. In fact, as a result of their repulsive interaction, the
distances between the $x_j$'s will grow with increasing length, until
eventually for sufficiently large $L$
\begin{equation}
x_1 \ll x_2 \ll \cdots \ll x_N.
\end{equation}
Then we may approximate
\begin{equation}
-{\partial\Omega\over\partial x_j} \approx N+1-2j,
\end{equation}
and find that Eq.\ (\ref{eq:FP bis}) is solved by a 
Gaussian distribution for the $x_j$,
\begin{eqnarray}
P(x_1,\cdots,x_N;L)&&\propto
  \exp
  \left\{
  \sum_{i,j=1}^N 
  -{\gamma\ell_{}\over4L} 
  \left(x_i-{L\over\xi_i}\right) 
  \right.
  \nonumber\\ 
  &&\hskip -40 pt 
  \times 
  \left.
  \left[\left(\openone_{N}-{1-\eta\over N}E_N\right)^{-1}\right]_{ij}
  \left(x_j-{L\over\xi_j}\right) 
  \vphantom{\sum_{i,j=1}^{N}} 
  \right\}.
  \label{eq: P for chiral in loc}
\end{eqnarray}
Here, the $N\times N$ matrix $E_N$ has the entries $(E_{N})_{ij}=1$, 
and  the channel-dependent ``localization length'' $|\xi_j|$ reads
\begin{equation}
\xi_j={\gamma\ell\over\beta(N+1-2j)}.
\end{equation}
For comparison, in the standard orthogonal and unitary symmetry
classes, the probability distribution $P(x_1,\cdots,x_N;L)$ in the
localized regime is also given by a Gaussian of the type (\ref{eq: P
for chiral in loc}), but with $\eta=1$, $\gamma = 2(\beta N + 2 -
\beta)$, and $\xi_j = (\beta N + 2 - \beta) \ell/(1 + \beta j -
\beta)$.\cite{BeenakkerReview}

In the localized regime $L \gg N \ell$ only the $x_j$ that are closest
to $0$ contribute to the conductance, cf.\ Eq.\ (\ref{eq:LandauerX}).
For even $N$, they are $x_{N/2}$ and $x_{(N/2)+1}$, 
both of which are an average distance
\begin{equation}
  \langle x_{N/2} \rangle = 
  -\langle x_{(N/2)+1} \rangle = {L\over\xi}, \qquad
  \xi = {\gamma \ell\over\beta}, 
\label{eq:xidef}
\end{equation}
away from zero. The length scale $\xi$ serves as the localization
length for even $N$. For odd $N$, the conductance is determined by
only one eigenvalue, $x_{(N+1)/2}$, which has zero average,
\begin{equation}
  \langle x_{(N+1)/2} \rangle = 0.
\end{equation}
The presence of the eigenvalue $x_{(N+1)/2}$ with zero average is
responsible for the absence of exponential localization in this case.

The average and variance of the conductance and the average and variance 
of its logarithm follow from the probability distribution 
(\ref{eq: P for chiral in loc}). 
For even $N$ the results are,
with an accuracy ${\cal O}(L^0/\xi^0)$ for the logarithms
displayed, 
\begin{mathletters}
\label{eq:chiral g}
\begin{eqnarray}
  \ln \langle g \rangle &=& 
  - {\beta\over 4} 
    \left(1-{1-\eta\over N}\right)^{-{1\over2}}{L\over\xi} 
  - {1\over2}\ln\left({L\over\xi}\right),
\label{eq:chiral_even_mean g} \\
  \ln \mbox{var}\, g &=& 
  \ln \langle g \rangle,
\label{eq:chiral_even_var g}
\end{eqnarray}
and
\begin{eqnarray}
  \langle \ln g \rangle &=&
  - {2 L \over \xi}  
  + 2\sqrt{{2\over\beta\pi}
    \left(1 - 2 {1 - \eta \over N} \right)
    {L\over\xi}},
\label{eq:chiral_even_mean lng}\\ 
  \mbox{var}\, \ln g &=&
  {4 \over \beta}
  \left[ 1 + \left( 1 - {2 \over \pi} \right) 
    \left(1 - 2 {1 - \eta \over N} \right) \right]{L \over\xi}.
\label{eq:chiral_even_var lng}
\end{eqnarray}
\end{mathletters}
The latter result shows that, in the localized regime, the conductance
distribution is well approximated by a log-normal distribution; 
unlike the average conductance $g$ itself,
which has fluctuations that are much bigger than the average, its
logarithm $\ln g$ provides a good characteristic of the ensemble.

For odd $N$, there is no exponential localization. The conductance has
a broad distribution, which is neither characterized by the (average of
the) conductance nor its logarithm,
\begin{equation}
  P(g) \propto
  {\exp \left[-{\gamma \ell \over 4 L} 
  \left( 1 - {1-\eta \over N} \right)^{-1}
  \mbox{arccosh}^2g^{-{1\over2}} \right] \over g\sqrt{1-g}}.
\end{equation}
With this distribution and up to corrections of order $L^0/\xi^0$, 
the average conductance decays algebraically,
\begin{mathletters} \label{eq:chiral g odd}
\begin{eqnarray}
\langle g \rangle&&= 
\left({\beta\over\pi}\right)^{1\over2}
\left(1 - {1-\eta\over N}\right)^{-{1\over2}}
\left({\xi\over L}\right)^{1\over2},
\label{eq:chiral_odd_mean g}
\\
\langle g^2\rangle&&= 
{2\over3}\langle g\rangle,
\label{eq:chiral_odd_mean gg}
\end{eqnarray} 
while the average of its logarithm grows
proportional to $L^{1/2}$ rather than $L$,
\begin{eqnarray} 
\langle \ln g \rangle&&= 
-4 
\sqrt{
{1\over\beta\pi}
\left(1 - {1-\eta\over N}\right)
{L\over\xi}},
\label{eq:chiral_odd_mean ln g}
\\
{\rm var}\, \ln g&&=
{8\over\beta}
\left(1-{2\over\pi}\right)
\left(1-{1-\eta\over N}\right)
{L\over\xi}.
\label{eq:chiral_odd_var ln g}
\end{eqnarray}
\end{mathletters}

Away from the band center $\varepsilon=0$, the conductance distribution
follows from the standard DMPK equation (\ref{eq:FP DMPK}).
It is close to log-normal,
with\cite{BeenakkerReview,Zirnbauer92,Mirlin94,BrouwerFrahm96}
\begin{mathletters}
\label{eq:std g}
\begin{eqnarray}
  \ln \langle g \rangle &=& -{L\over2\xi_{\rm st}}
    - {3 \over 2} \ln \left( {L\over\xi_{\rm st}}\right), 
\label{eq:std g a}\\
  \ln \mbox{var}\, g &=& \ln \langle g \rangle, 
\label{eq:std g b}\\
  \langle \ln g \rangle &=& 
     - {2 L\over\xi_{\rm st}}, 
\label{eq:std g c}\\
  \mbox{var}\, \ln g &=& 
     {4L\over\xi_{\rm st}},
\label{eq:std g d}
\end{eqnarray}
\end{mathletters}%
up to an accuracy of ${\cal O}(L^0/\xi^0)$.
Here the localization length for the standard symmetry classes is given
by
\begin{equation}
  \xi_{\rm st} = (\beta N + 2 - \beta)\ell.
  \label{eq:loc length standard}
\end{equation}

The most striking difference in the conductance distribution appears
for odd $N$, where the absence of exponential
localization at $\varepsilon=0$ is contrasted with the exponential
decay of the conductance for $\varepsilon \neq 0$.
However, also for an even number of channels,
there is an important difference. At $\varepsilon=0$, the localization
length $\xi\approx N\ell$ is $\beta$-independent for large $N$,
cf.\ Eqs.\ (\ref{eq:gamma}) and (\ref{eq:xidef}), 
while the localization length $\xi_{\rm st}\approx\beta N\ell$ 
away from the band center is proportional to $\beta$ for
$\varepsilon \neq 0$, cf.\ Eq.\ (\ref{eq:loc length standard}). 
Hence, upon
moving away from the band center, the localization length increases by
a factor $\beta$. [The mean free path does not depend on $\varepsilon$,
see Eq.\ (\ref{eq:Born}).]


The absence of a $\beta$-dependence for the localization length at the
band center may be related to the anomaly in the DoS for
random hopping models at that energy. In Ref.\ \onlinecite{Brouwer99DOS}, 
it was shown that in the absence of time-reversal symmetry the DoS
$\rho(\varepsilon)$ near zero energy has a pseudogap,
$\rho(\varepsilon) \propto \varepsilon |\ln \varepsilon|$, while in
the presence of time-reversal symmetry $\rho$ has a logarithmic
divergence, $\rho(\varepsilon) \propto |\ln \varepsilon|$. We conclude
that, upon breaking time-reversal symmetry, the decrease in the DoS
available for transport, cancels the suppression of
destructive interference responsible for the increase of the
localization length in the standard case.

The average and variance of the conductance in the localized regime are
dominated by rare events, where the smallest $x_j$ is close to zero
(corresponding to a transmission coefficient close to unity).  For
wires without chiral symmetry, approximation of $P(x_1,\ldots,x_N;L)$ by
a Gaussian similar to Eq.\ (\ref{eq: P for chiral in loc}) fails for
$x_j$ close to zero because it does not account for the repulsion
between $x_j$ and its mirror image $-x_j$, cf.\ Eq.\ (\ref{eq:FP
DMPK}). While it does not affect the leading ${\cal O}(L)$ behavior
of $\ln \langle g \rangle$ and $\ln \mbox{var}\, g$, this failure shows
up in the subleading logarithmic terms in 
Eqs.\ (\ref{eq:std g a},\ref{eq:std g b}) 
which are different from what one would have obtained from a Gaussian
distribution for the $x_j$. [The results quoted in
Eqs.\ (\ref{eq:std g a},\ref{eq:std g b}) 
above follow from an exact solution of the DMPK
equation.\cite{Zirnbauer92,Mirlin94,BrouwerFrahm96}] In the presence
of the chiral symmetry, there is no repulsion between $x_j$ and $-x_j$,
so that the approximation (\ref{eq: P for chiral in loc}) remains valid
for $x_j$ close to zero. In this respect, we remark that the
logarithmic terms in Eq.\ (\ref{eq:chiral g}), which were obtained
with the help of Eq.\ (\ref{eq: P for chiral in loc}) indeed agree with
the exact solution of Ref.\ \onlinecite{Mudry99} for the case
$\beta=2$.

\section{Diffusive regime}
\label{sec:Diffusive regime}

In the diffusive regime $\ell \ll L \ll N \ell$, the effects of quantum
interference do not take such a dramatic form as in the localized
regime. The typical conductance of any sample is given by the classical
Ohm's law, $g = N \ell/L$, and does not know of quantum mechanical
phase coherence, the presence or absence of time-reversal symmetry, or,
as we shall see below, the presence or absence of chiral symmetry.  The
role of quantum mechanics, and hence the role of the symmetries of the
microscopic Hamiltonian in this regime is confined to small corrections to 
the average conductance and to its sample-to-sample fluctuations. In spite
of their smallness, these corrections are of prime importance, as they
are a universal signature of quantum phase coherence, their size being
determined  by the fundamental symmetries of the system only. They do
not depend on microscopic properties of the quantum wire, nor on its
macroscopic characteristics, such as mean free path, width, or length.

The two corrections are referred to as ``weak-localization'' and
``universal conductance fluctuations''.  The former is a small
correction $\delta g$ ($\delta p$) to the ensemble averaged (dimensionless)
conductance $\langle g \rangle$ (shot-noise $\langle p\rangle$)
that is suppressed if time-reversal
symmetry is broken by a magnetic field. For a standard quantum wire, it 
reads
\begin{equation}
  \delta g = {\beta - 2 \over3\beta}
\qquad
  \left(\delta p = {\beta-2\over45\beta}\right).
\end{equation}
Since it signals the first departure from Ohm's law, the weak
localization correction to the conductance is precursor to the
exponential suppression of the conductance in the localized regime.
The universal conductance fluctuations refer to the sample-to-sample
fluctuations of the conductance, which have variance,
\begin{equation}
  \mbox{var}\, g = {2 \over 15 \beta}. \label{eq:var g st}
\end{equation}  
The breaking of time-reversal (spin-rotation) symmetry reduces the
conductance fluctuations by a universal factor of $\sqrt{2}$ ($2$).

In this section we calculate those quantum corrections for the case of
a quantum wire with random hopping. 
Our calculations are inspired by the approach
of Mello and Stone,\cite{Mello91} who have derived and solved scaling
equations for the moments of the conductance in the standard
universality classes from the DMPK equation in the limit of large $N$.
We consider both the quantum corrections for the pure symmetry classes,
corresponding to the Fokker-Planck equations (\ref{eq:FP ch}) and
(\ref{eq:FP DMPK}) at $\varepsilon=0$ and $\varepsilon$ far away from
$0$, respectively, and for the intermediate regime, where the crossover
between the two symmetry classes takes place. Since in the latter case
no Fokker-Planck equation for the transmission eigenvalues $x_j$ is
available, a modification of the approach of Ref.\ \onlinecite{Mello91}
is needed, which is based on the more fundamental scaling equation for the
reflection matrix $r$, Eq.\ (\ref{eq: r to r + delta r}), rather than
on a Fokker-Planck equation for the transmission
eigenvalues $x_j$. Such a method was proposed by one of the authors
\cite{Brouwer98WAV} in the context of the transmission through a random
waveguide with absorption. Below, we adapt this method to the present
case (Sec.\ \ref{sec:RG equations}), and present solutions for the chiral
symmetry classes at the band center $\varepsilon=0$
(Sec.\ \ref{sec:Diffusive regime in the chiral limit}) and for the
crossover from the chiral symmetry classes to the standard universality
classes as $\varepsilon$ moves away from the band center
$\varepsilon=0$ (Sec.\ \ref{sec:Crossover between the chiral and
standard universality classes}).


\subsection{Scaling equations}
\label{sec:RG equations}

Although we are primarily interested in the statistics of the
transmission matrix $t$, and in particular in the (dimensionless)
conductance $g = \mbox{tr}\, t^{\dagger} t$ and shot noise power
$p = \mbox{tr}\left[ t^{\dagger} t( {1} - t^{\dagger} t)\right]$,
we find it more convenient to formulate our scaling equations in terms of
the reflection matrix $r$. Once we know $r$, unitary of the scattering
matrix allows us to find the transmission properties without much
effort.

Before we write down the most general scaling equation for a trace of an
arbitrary product of the reflection matrix $r$ and its Hermitian
conjugate, we would like to focus on the scaling equation for $\mbox{tr}\,
r^{\dagger} r$ in order to demonstrate the method and the
approximations involved. 
Addition of a thin disordered slice to a disordered wire causes a small
change $r \to r + \delta r$ to the reflection matrix $r$, see
Eq.\ (\ref{eq:delta r}). Hence, upon addition of this slice, the trace
$\mbox{tr}\, r^{\dagger} r$ changes to $\mbox{tr}\, (r + \delta
r)^{\dagger} (r + \delta r)$. Using Eq.\ (\ref{eq:delta r}) for $\delta
r$, we thus find, up to ${\cal O}(\delta L)$ [Recall that the variance
of $W$ is of order $\delta L$, so keeping terms up to ${\cal O}(\delta
L)$ means up to ${\cal O}(W^2)$],
\bleq 
\begin{eqnarray}
\delta\,{\rm tr}\,r^{\dag}r&=&
  - \mbox{tr}\,
  \big[r ({1}-r^{\dag} r) W
   + W ({1} - r^{\dag} r) r^{\dag}
\nonumber\\
&&
 - ({1} - rr^{\dag}) W ({1} - r^{\dag} r) W
 +r ({1} - r^{\dag} r)WrW
 + Wr^{\dag}W({1} - r^{\dag} r) r^{\dag}
\big].
\label{eq: rdag r appr}
\end{eqnarray}
All terms that involve the disorder potential $V$ in Eq.\
(\ref{eq:delta r}) canceled due to the cyclicity of the trace.
Next we perform a disorder 
average over $W$ and over the reflection matrix $r$ of the wire of
length $L$. We thus find
\begin{eqnarray}
  {\gamma \ell \over \beta}
  {\delta\, \langle \mbox{tr}\, r^{\dag}r \rangle \over \delta L}
  &=& \langle [\mbox{tr}\, ({1} - r^{\dag} r)]^2 \rangle
  - \langle \mbox{tr}\, r\, \mbox{tr}\, r ({1} - r^{\dag} r) \rangle
  - \langle \mbox{tr}\, r^{\dag}\,
      \mbox{tr}\, r^{\dag} ({1} - r r^{\dag}) \rangle
\nonumber\\
&&
  \mbox{} + {{2-\beta\over\beta}} \langle
  \mbox{tr}\, 
({1} - r^{\dag} r)
({1}- r^{\dag} r - r^2 - r^{\dag2}) \rangle
  \mbox{} -{2\left(1-\eta\right)\over\beta N}
  \langle
  \mbox{tr}\, ({1} - r r^{\dag})
              ({1} - r^{\dag} r - r^2 - r^{\dag2}) \rangle.
\label{eq:delta trace r dag r}
\end{eqnarray}
\eleq 
\noindent
Finally, we take the limit $\delta L \ll \ell$, and replace the finite
differences on the l.h.s.\ of Eq.\ (\ref{eq:delta trace r dag r}) by
differentials.

It is apparent that the scaling equation obeyed by $\langle{\rm tr}\,
r^{\dag}r\rangle$ is not closed: On the r.h.s.\ traces and
products of traces of up to four reflection matrices appear. Closure requires
an infinite family of scaling equations, and cannot be achieved on the level
of scaling equations for the moments, but only with the help of the
Fokker-Planck equation for the transmission eigenvalues $x_j$ in the
cases of pure symmetry. However, for lengths $L \ll N \ell$ it is
possible to decouple this infinite set, and to find a solution order by
order in $L/(N\ell)$. Formally, this decoupling scheme proceeds along
the lines of a large-$N$ expansion: In addition to the explicit factors
$N$ in Eq.\ (\ref{eq:delta trace r dag r}), each trace contributes a
factor $N$. Further, we assume that, to leading order in $N$, the
average of a product of traces equals the product of the averages. As
we will see below, corrections correspond to a (co)variance of traces,
and are of order $N^{0}$. Similarly, if we have a product of $n$
traces, we can expand in cumulants, where an $n$th cumulant will turn
out to be of relative size $N^{2-n}$. Such a decoupling scheme is known
to work for the case of the standard DMPK equation,\cite{Mello91} and
its consistency can be verified from the scaling equations for traces and
products of traces that we derive in this section.

Let us now see how the scaling equation for $\langle \mbox{tr}\, r^{\dagger}
r \rangle$ decouples in this large-$N$ decoupling scheme.
Recalling that $\gamma$ is of order $N$, cf.\ Eq.\ (\ref{eq:gamma}),
we thus find that the r.h.s.\ of Eq.\ (\ref{eq:delta trace r dag r})
is of order $N^2$, i.e.,
\begin{equation}
  \ell \partial_L \left\langle {\rm tr}\, r^{\dag}r \right\rangle =
 N
-2\left\langle {\rm tr}\, r^{\dag}r \right\rangle
+{1\over N}\left\langle {\rm tr}\, r^{\dag}r \right\rangle^2
+{\cal O}(N^0).
\label{eq: leading tra r dag r}
\end{equation}
Here we have used the fact that the average of the trace of an odd
product of $r$'s and $r^{\dag}$'s is zero, see our discussion below
Eq.\ (\ref{eq: r to r + delta r}).
The initial condition at $L=0$ corresponds to
perfect transmission, i.e., $\langle {\rm tr}\, r^{\dag}r \rangle=0$.
The solution is easily found,
\begin{equation}
\left\langle {\rm tr}\, r^{\dag}r \right\rangle=
{Ns\over s+1}+{\cal O}(N^0),
\end{equation}
where $s = L/\ell$. This solution corresponds to Ohm's law for the
conductance $g = N - \mbox{tr}\, r^{\dagger} r$,
\begin{equation}
 \langle g \rangle = {N\over s+1}+{\cal O}(N^0).
\label{eq: example for tr rdag r to leading order}
\end{equation}

To this order in $N$, the result is entirely classical. The average
$\langle {\rm tr}\, r^{\dag}r \rangle$ (and hence $\langle g \rangle$)
does not depend on the energy $\varepsilon$ nor on the presence or
absence of time-reversal symmetry. The dependence on time-reversal
symmetry shows up through the term proportional to ${{2-\beta\over\beta}}$
on the r.h.s.~of Eq.\ (\ref{eq:delta trace r dag r}), 
which is of order $N$. It is
this term in the scaling equation that gives rise to the weak-localization
correction to the conductance. The scaling equation for $\langle {\rm tr}\,
r^{\dag}r \rangle$ does not contain an explicit energy dependence.
Instead, the energy-dependence shows up through the appearance of the
traces like $\mbox{tr}\, r^2$ or $\mbox{tr}\, r^{\dagger} r^3$ in
Eq.\ (\ref{eq:delta trace r dag r}) in the weak-localization
correction. Such traces that contain different numbers of $r$'s and
$r^{\dagger}$'s strongly depend on energy, as can be seen from the scaling
equation of, e.g., $\langle \mbox{tr}\, r^2 \rangle$,
\bleq 
\begin{equation}
  {\gamma \ell \over \beta} \partial_L \langle{\rm tr}\, r^2\rangle =
  \langle[\mbox{tr}\, ({1} - r^2)]^2\rangle - 
  2 \langle \mbox{tr}\, r\, \mbox{tr}\, r({1}-r^2) \rangle
  + {4 {i} \varepsilon \gamma \ell \over \beta} 
  \langle {\rm tr}\, r^2 \rangle
  \mbox{} + \left[{{2-\beta\over\beta}} -{2(1-\eta)\over\beta N}\right]
  \langle {\rm tr}\, ({1} - r^2)({1} - 3 r^2)  \rangle.
\label{eq: tr rr to leading order}
\end{equation}
With the same decoupling scheme as before, we find a closed scaling equation
for $\langle{\rm tr}\, r^2\rangle$ up to ${\cal O}(N)$,
\begin{equation}
  \ell \partial_L \langle{\rm tr}\, r^2\rangle =
  N - 2 (1- 2 {i} \varepsilon \ell) 
\langle{\rm tr}\, r^2\rangle 
  + {1\over N} \langle{\rm tr}\, r^2\rangle^2 + {\cal O}(N^0),
\end{equation}
which has the solution
\begin{eqnarray}
  \left\langle {\rm tr}\, r^2 \right\rangle
  &=&
  N 
  \left\{ 1 - 2 {i} \varepsilon \ell + 
  2\sqrt{\varepsilon \ell({i}+\varepsilon \ell)}
  \cot\! \left[ 2\sqrt{\varepsilon \ell({i}+\varepsilon \ell)}\, s
       \right] 
  \right\}^{-1} +{\cal O}(N^0).
\label{eq: cross tr rr leading}
\end{eqnarray}
One verifies that for $\varepsilon \to 0$, the average $\langle
\mbox{tr}\, r^2 \rangle$ equals the average $\langle \mbox{tr}\,
r^{\dag} r \rangle$ that we computed above, since for zero energy one
has $r = r^{\dagger}$.
One also verifies that for $\varepsilon\ell \gg 1$
the average $\langle \mbox{tr}\, r^2 \rangle$ approaches zero, as is
the case in the standard symmetry classes. 

We are now ready to discuss the scaling equations for the trace of the
product of an arbitrary number 
of reflection matrices and for the product of such traces.
Hereto we write $r_0 = r$ and $r_1 = r^{\dagger}$, and define
\begin{equation}
R_{j_{1}\cdots j_{n}} \equiv {\rm tr}\ r_{j_{1}}\ldots r_{j_{n}},
\label{eq:def for R}
\end{equation}
where the indices $j_k$ can take the values $0$ or $1$. We define the
symbol $R$ without indices as $R=N$. We also define products of
traces through the symbols
\begin{equation}
Q^{\ }_{{\bf n}_1\cdots {\bf n}_m}=
R^{\ }_{i^{(1)}_1\cdots i^{(1)}_{n_1}}
\ldots
R^{\ }_{i^{(m)}_1\cdots i^{(m)}_{n_m}},
\end{equation}
where ${\bf n}_j$ denotes the $n$-tuple $i^{(j)}_1,\ldots,i^{(j)}_{n_j}$.

Proceeding along the same lines as above, we then find that the scaling
equation for a single trace is given by (see Fig.\ \ref{fig:diagramatic})
\begin{eqnarray}
{\gamma \ell\over\beta}\partial_L&&
  \left\langle R_{j_{1}\cdots j_{n}}\right\rangle= 
  {2{i} \varepsilon \gamma \ell [\sum_{k=1}^n(-1)^{j_k}] - n\gamma \over \beta}
  \left\langle R_{j_{1}\cdots j_{n}}\right\rangle
\nonumber\\ 
&&
  + \sum_{1\leq k\leq l\leq n} \hskip -6 pt
  \left\langle
  R_{j_{k  }\cdots j_{l  }} R_{j_{l  }\cdots j_{n}j_{1}\cdots j_{k  }} 
  + {{2-\beta\over\beta}}
  R_{j_{k  }\cdots j_{l  }j_{k }\cdots j_{1}j_{n}\cdots j_{l  }}
  -\frac{2(1-\eta)}{\beta N}
  R_{j_{k  }\cdots j_{l  }j_{l }\cdots j_{n}j_{1}\cdots j_{k  }}
  \right \rangle
\nonumber\\ 
&&
  + \sum_{1\leq k<l\leq n} \hskip -6 pt
  \left\langle
  R_{j_{k+1}\cdots j_{l-1}} R_{j_{l+1}\cdots j_{n}j_{1}\cdots j_{k-1}}
  + {{2-\beta\over\beta}}
  R_{j_{k+1}\cdots j_{l-1}j_{k-1}\cdots j_{1}j_{n}\cdots j_{l+1}}
  -\frac{2(1-\eta)}{\beta N}
  R_{j_{k+1}\cdots j_{l-1}j_{l+1}\cdots j_{n}j_{1}\cdots j_{k-1}}
  \right \rangle 
\nonumber\\
&&
  - \sum_{1\leq k<l\leq n} \hskip -6 pt
  \left\langle
  R_{j_{k  }\cdots j_{l-1}} R_{j_{l+1}\cdots j_{n}j_{1}\cdots j_{k  }} 
  + {{2-\beta\over\beta}}
  R_{j_{k  }\cdots j_{l-1}j_{k  }\cdots j_{1}j_{n}\cdots j_{l+1}} 
  -\frac{2(1-\eta)}{\beta N}
  R_{j_{k  }\cdots j_{l-1}j_{l+1}\cdots j_{n}j_{1}\cdots j_{k  }} 
  \right\rangle 
\nonumber\\ 
&&
  - \sum_{1\leq k<l\leq n}\hskip -6 pt
  \left\langle
  R_{j_{k+1}\cdots j_{l  }} R_{j_{l  }\cdots j_{n}j_{1}\cdots j_{k-1}}
  + {{2-\beta\over\beta}}
  R_{j_{k+1}\cdots j_{l  }j_{k-1}\cdots j_{1}j_{n}\cdots j_{l  }}
  -\frac{2(1-\eta)}{\beta N}
  R_{j_{k+1}\cdots j_{l  }j_{l  }\cdots j_{n}j_{1}\cdots j_{k-1}}
  \right\rangle. 
  \label{eq: incre any one traces of r's}
\end{eqnarray}
Here, it is understood that 
$j_{n+1}\equiv j_1$, $j_{0}\equiv j_n$.
Moreover for $n\geq l=k+1>1$,
$
R_{j_{k+1}\cdots j_{l-1}}\equiv{\rm tr}\,\openone_N=N
$,
$
R_{j_{k+1}\cdots j_{l-1}j_{k-1}\cdots j_1j_n\cdots j_{l+1}}\equiv
R_{j_{k-1}\cdots j_1j_n\cdots j_{k+2}}
$,
and
$
R_{j_{k+1}\cdots j_{l-1}j_{l+1}\cdots j_nj_1\cdots j_{k-1}}\equiv
R_{j_{k+2}\cdots j_nj_1\cdots j_{k-1}},
$
respectively, 
whereas when $k=1$ and $l=n$
$
R_{j_{l+1}\cdots j_nj_1\cdots j_{k-1}}\equiv{\rm  tr}\, \openone_N=N,
$
$
R_{j_{k+1}\cdots j_{l-1}j_{k-1}\cdots j_{1}j_{n}\cdots j_{l+1}}\equiv
R_{j_{k+1}\cdots j_{l-1}j_{l+1}\cdots j_{n}j_{1}\cdots j_{k-1}}\equiv
R_{j_2\cdots j_{n-1}},
$
respectively.
Note that there is a one-to-one correspondence between
contributions involving a product of two traces, say,
$$
  R_{j_{k  }\cdots j_{l  }} R_{j_{l  }\cdots j_{n}j_{1}\cdots j_{k  }}\equiv
{\rm tr}\,
\left(
r_{j_{k  }}\cdots r_{j_{l  }}
\right)
{\rm tr}\,
\left(
r_{j_{l  }}\cdots r_{j_{n}}r_{j_{1}}\cdots r_{j_{k  }}
\right),
$$
and contributions arising in the presence of time reversal symmetry
$$
{2-\beta\over\beta}
  R_{j_{k  }\cdots j_{l  }j_{k  }\cdots j_{1}j_{n}\cdots j_{l  }}\equiv
{2-\beta\over\beta}
{\rm tr}\,
\left[
\left(
r_{j_{k  }}\cdots r_{j_{l  }}
\right)
\left(
r_{j_{k  }}\cdots r_{j_{1}}r_{j_{n}}\cdots r_{j_{l  }}
\right)
\right]=
{2-\beta\over\beta}
{\rm tr}\,
\left[
\left(
r_{j_{k  }}\cdots r_{j_{l  }}
\right)
\left(
r_{j_{l  }}\cdots r_{j_{n}}r_{j_{1}}\cdots r_{j_{k  }}
\right)^{\rm t}
\right],
$$
or due to the randomness in the determinant of the hopping matrices
$$
  -\frac{2(1-\eta)}{\beta N}
  R_{j_{k  }\cdots j_{l  }j_{l }\cdots j_{n}j_{1}\cdots j_{k  }}\equiv
{\rm tr}\,
\left[
\left(
r_{j_{k  }}\cdots r_{j_{l  }}
\right)
\left(
r_{j_{l  }}\cdots r_{j_{n}}r_{j_{1}}\cdots r_{j_{k  }}
\right)
\right].
$$
For products of traces, we find
\begin{mathletters}
\label{eq: scaling Q with F}
\begin{eqnarray}
  {\gamma \ell \over \beta} \partial_L
  \left\langle
  Q_{{\bf n}_1\cdots {\bf n}_m}
  \right\rangle
  &=&
  \sum_{j=1}^m
  \left\langle {\gamma \ell \over \beta} 
  Q_{{\bf n}_{1  }\cdots {\bf n}_{j-1}{\bf n}_{j+1}\cdots {\bf n}_{m  }}
  \partial_L R_{{\bf n}_{j}} \right\rangle
  + \sum_{1\leq k<l\leq m}
  \left\langle
  Q^{\ }_{{\bf n}_{1  }\cdots {\bf n}_{k-1}
          {\bf n}_{k+1}\cdots {\bf n}_{l-1}
          {\bf n}_{l+1}\cdots {\bf n}_{m  }}
  F^{\ }_{{\bf n}_k{\bf n}_l}
  \right\rangle,
\label{eq: scaling for Q}
\end{eqnarray}
where $(\gamma \ell/\beta)\partial_L R_{{\bf n}_j}$ is given by the
r.h.s.\ of Eq. (\ref{eq: incre any one traces of r's}) with omission
of the angular brackets for the disorder averaging, and where
$
F^{\ }_{{\bf mn}}=
\sum_{k=1}^m\sum_{l=1}^n f_{k,l}
$
with
\begin{eqnarray}
f_{k,l}&&=
 R^{\ }_{i_{k  }\cdots i_{m  }i_{1  }\cdots i_{k  } 
         j_{l  }\cdots j_{n  }j_{1  }\cdots j_{l  }}
+{{2-\beta\over\beta}}
 R^{\ }_{i_{k  }\cdots i_{m  }i_{1  }\cdots i_{k  } 
         j_{l  }\cdots j_{1  }j_{n  }\cdots j_{l  }}
-\frac{2(1-\eta)}{\beta N}
 R^{\ }_{i_{k  }\cdots i_{m  }i_{1  }\cdots i_{k  }} 
 R^{\ }_{j_{l  }\cdots j_{n  }j_{1  }\cdots j_{l  }}
\nonumber\\
&&
+
 R^{\ }_{i_{k+1}\cdots i_{m  }i_{1  }\cdots i_{k-1} 
         j_{l+1}\cdots j_{n  }j_{1  }\cdots j_{l-1}}
+{{2-\beta\over\beta}}
 R^{\ }_{i_{k+1}\cdots i_{m  }i_{1  }\cdots i_{k-1} 
         j_{l-1}\cdots j_{1  }j_{n  }\cdots j_{l+1}}
-\frac{2(1-\eta)}{\beta N}
 R^{\ }_{i_{k+1}\cdots i_{m  }i_{1  }\cdots i_{k-1}} 
 R^{\ }_{j_{l+1}\cdots j_{n  }j_{1  }\cdots j_{l-1}}
\nonumber\\
&&
-
 R^{\ }_{i_{k  }\cdots i_{m  }i_{1  }\cdots i_{k  } 
         j_{l+1}\cdots j_{n  }j_{1  }\cdots j_{l-1}}
-{{2-\beta\over\beta}}
 R^{\ }_{i_{k  }\cdots i_{m  }i_{1  }\cdots i_{k  } 
         j_{l-1}\cdots j_{1  }j_{n  }\cdots j_{l+1}}
+\frac{2(1-\eta)}{\beta N}
 R^{\ }_{i_{k  }\cdots i_{m  }i_{1  }\cdots i_{k  }} 
 R^{\ }_{j_{l+1}\cdots j_{n  }j_{1  }\cdots j_{l-1}}
\nonumber\\
&&
-
 R^{\ }_{i_{k+1}\cdots i_{m  }i_{1  }\cdots i_{k-1} 
         j_{l  }\cdots j_{n  }j_{1  }\cdots j_{l  }}
-{{2-\beta\over\beta}}
 R^{\ }_{i_{k+1}\cdots i_{m  }i_{1  }\cdots i_{k-1} 
         j_{l  }\cdots j_{1  }j_{n  }\cdots j_{l  }}
+\frac{2(1-\eta)}{\beta N}
 R^{\ }_{i_{k+1}\cdots i_{m  }i_{1  }\cdots i_{k-1}} 
 R^{\ }_{j_{l  }\cdots j_{n  }j_{1  }\cdots j_{l  }}.
\nonumber\\
\label{eq:Fmn}
\end{eqnarray}
\end{mathletters}
\eleq 
\noindent
Here we denoted ${\bf m}=i_1,\ldots,i_m$ and ${\bf n}=j_1,\ldots,j_n$.

Below, we are interested in averages and (co)variances of traces of an
even number of reflection matrices up to order $N^0$. In both cases,
the terms proportional to $(1-\eta)$ do not play a role. For the
average of a single trace, this is immediately clear from
Eq.\ (\ref{eq: incre any one traces of r's}). To see this for the
(co)variance of two traces, some further inspection of 
Eq.\ (\ref{eq: scaling Q with F}) 
is needed. First, $\eta$ appears explicitly in the
quantity $F_{{\bf mn}}$, multiplying a product of two traces, see
Eq.\ (\ref{eq:Fmn}). A priori, the leading contribution, which is
obtained by replacement of those traces by their averages, is of the
same order [${\cal O}(N)$] as the other terms in Eq.\ (\ref{eq:Fmn}).
However, as $m$ and $n$ are even, each of the two traces multiplying
$(1-\eta)$ contains an odd number of reflection matrices, so that their
averages vanish. Hence, to leading order in $N$, the contribution from
the term proportional to $(1-\eta)$ vanishes. Second, $\eta$ appears
implicitly through the derivative $(\gamma \ell/\beta)\partial_L
R_{{\bf n}_j}$ in Eq.\ (\ref{eq: scaling for Q}).  Again, to leading order
in $N$, its contribution vanishes, and one is left with a term of
relative size $N^{-2}$.

It should be mentioned that 
Eqs.~(\ref{eq: incre any one traces of r's}) 
and (\ref{eq: scaling Q with F})
can be extended to the case in which a weak 
staggering of the hopping amplitude is present in the microscopic model, 
cf.~Eqs.~(\ref{eq:Schr eq}--\ref{eq:gamma}). 
(How to generalize the Fokker-Planck equation (\ref{eq:FP ch})
to include dimerization
was shown in Ref.~\onlinecite{Brouwer98}, see also 
Ref.~\onlinecite{Brouwer99DOS}.)
Weak staggering of the hopping amplitude is implemented by requiring
that the disorder potential $W$ has the Gaussian distribution with 
variance (\ref{eq: var W}) and mean 
$\langle W_{jk}\rangle={\beta\delta L\over\gamma\ell}\Delta\delta_{jk}$.
Here $\Delta$ measures the strength of the dimerization along the chain
direction. With weak dimerization,
Eq.~(\ref{eq: incre any one traces of r's}), 
say, is modified by the addition on the r.h.s.~of the contribution
$
\Delta\sum_{k=1}^n
\langle 
R_{j_k\cdots j_n\cdots j_k}
-
R_{j_{k+1}\cdots j_n\cdots j_{k-1}}
\rangle
$.
We see that the scaling equations now couple traces over an even and odd
number of reflection matrices as is expected since the probability
distribution of $W$ is not anymore symmetric about $W=0$,
cf.~Eq.~(\ref{eq:delta r}).

Equations (\ref{eq: incre any one traces of r's}) and (\ref{eq: scaling Q
with F}) are the central results of this section. These equations are
more general than the Fokker-Planck equations (\ref{eq:FP ch}) and
(\ref{eq:FP DMPK}) in the sense that they are valid both at the center of
the band $\varepsilon=0$ and in its proximity.  Their limitation
is that they can only be solved in the diffusive regime $L \ll N \ell$.
In particular, they cannot be used to probe the localized regime (in
contrast to their counterparts in the problem of a wave guide with
absorption, see Ref.\ \onlinecite{Brouwer98WAV}).
The next two subsections are devoted to a solution in the diffusive regime. 
The case of pure chiral symmetry ($\varepsilon=0$) is considered 
in Sec.\ \ref{sec:Diffusive regime in the chiral limit}; 
the energy dependence of the solution is 
discussed in 
Sec.\ \ref{sec:Crossover between the chiral and standard universality classes}.

\begin{figure}
\begin{center}
\setlength{\unitlength}{0.75mm}
\begin{picture}(90,105)(5,-50)
\thicklines

\put( -6, 5){\line(1, 0){36}}
\put( -6, 5){\line(0,-1){10}}
\multiput(0,0)(6, 0){5}{\circle{2}} 
\put(30,-5){\line(0, 1){10}}
\put( -6,-5){\line(1,0){36}} 
\put(32,-1){$\longrightarrow$}


\put( 82, 44){$(\hbox{a})$}

\put( 44, 50){\line(1, 0){36}}
\put( 44, 50){\line(0,-1){10}}
\multiput(50,45)(6, 0){5}{\circle{2}} 
\put(80,40){\line(0, 1){10}}
\put( 44,40){\line(1,0){36}}


\put( 82, 29){$(\hbox{b})$}

\put( 44, 36){\line( 1, 0){22}}\put( 44, 24){\line( 1, 0){22}}
\put( 44, 36){\line( 1, 0){22}}\put( 44, 24){\line( 1, 0){22}}
\put( 80, 35){\line(-1, 0){22}}\put( 80, 25){\line(-1, 0){22}}
\put( 66, 36){\line( 0,-1){12}}\put( 58, 35){\line( 0,-1){10}}
\put( 60, 34){\line(0,-1){ 8}}\put( 64, 34){\line(0,-1){ 8}}
\put( 60, 34){\line(1, 0){ 4}}\put( 60, 26){\line(1, 0){ 4}}
\multiput(50,30)(6, 0){5}{\circle{2}}\put(62,30){\circle*{2}} 


\put( 82, 14){$(\hbox{b'})$}

\put( 44, 20){\line( 1, 0){15}}\put( 80, 20){\line(-1, 0){15}}
\put( 44, 10){\line( 1, 0){15}}\put( 80, 10){\line(-1, 0){15}}
\put( 59, 20){\line( 0,-1){10}}\put( 65, 20){\line( 0,-1){10}}
\put( 54, 19){\line( 0,-1){ 8}}\put( 70, 19){\line( 0,-1){ 8}}
\put( 54, 19){\line( 1, 0){16}}\put( 54, 11){\line( 1, 0){16}}
\multiput(50, 15)(6, 0){5}{\circle{2}}\multiput( 56, 15)(12,0){2}{\circle*{2}} 


\put( 82, -1){$(\hbox{c})$}

\put( 44,  5){\line( 1, 0){ 9}}\put( 80,  5){\line(-1, 0){ 9}}
\put( 44, -5){\line( 1, 0){ 9}}\put( 80, -5){\line(-1, 0){ 9}}
\put( 53,  5){\line( 0,-1){10}}\put( 71,  5){\line( 0,-1){10}}
\put( 60, +4){\line( 0,-1){ 8}}\put( 64,  4){\line( 0,-1){ 8}}
\put( 60, +4){\line( 1, 0){ 4}}\put( 60, -4){\line( 1, 0){ 4}}
\multiput(50,  0)(6, 0){5}{\circle{2}}\multiput( 56,  0)(12,0){2}{\circle*{2}} 


\put( 82,-31){$(\hbox{e})$}

\put( 44,-10){\line( 1, 0){15}}\put( 80,-10){\line(-1, 0){ 9}}
\put( 44,-20){\line( 1, 0){15}}\put( 80,-20){\line(-1, 0){ 9}}
\put( 59,-10){\line( 0,-1){10}}\put( 71,-10){\line( 0,-1){10}}
\put( 54,-11){\line( 0,-1){ 8}}\put( 64,-11){\line( 0,-1){ 8}}
\put( 54,-11){\line( 1, 0){10}}\put( 64,-19){\line(-1, 0){10}}
\multiput(50,-15)(6, 0){5}{\circle{2}}\multiput( 56,-15)(12,0){2}{\circle*{2}} 

%


\put( 82,-16){$(\hbox{d})$}

\put( 44,-25){\line( 1, 0){ 9}}\put( 80,-25){\line(-1, 0){15}}
\put( 44,-35){\line( 1, 0){ 9}}\put( 80,-35){\line(-1, 0){15}}
\put( 53,-25){\line( 0,-1){10}}\put( 65,-25){\line( 0,-1){10}}
\put( 60,-26){\line( 0,-1){ 8}}\put( 70,-26){\line( 0,-1){ 8}}
\put( 60,-26){\line( 1, 0){10}}\put( 70,-34){\line(-1, 0){10}}
\multiput(50,-30)(6, 0){5}{\circle{2}}\multiput( 56,-30)(12,0){2}{\circle*{2}}

\put(40,-45){$\quad +\cdots $}

%
%
\end{picture}
\end{center}
\caption{
Diagrammatic representation of 
Eq.~(\ref{eq: incre any one traces of r's}). Each circle corresponds
to a reflection matrix $r$ or $r^{\dag}$. To calculate the increment of
$\langle R_{j_1\cdots j_n}\rangle$ (single box containing $n$ open circles)
one chooses a pair of (filled) circles. As indicated in (b) the same
circle can be chosen twice. 
Overlapping or nested boxes represent multiplication of traces. Thus, 
$\langle R_{j_1j_2j_3j_4j_5}\rangle$,
$\langle R_{j_3}R_{j_3j_4j_5j_1j_2j_3}\rangle$,
$\langle R_{j_2j_3j_4}R_{j_4j_5j_1j_2}\rangle$,
$\langle R_{j_3}R_{j_5j_1}\rangle$,
$\langle R_{j_2j_3}R_{j_5j_1j_2}\rangle$, and
$\langle R_{j_3j_4}R_{j_4j_5j_1}\rangle$, 
are represented by (a), (b), (b'), (c), (d), and (e), respectively.
}
\label{fig:diagramatic}
\end{figure}

\subsection{Diffusive regime in the chiral limit}
\label{sec:Diffusive regime in the chiral limit}

The general scaling equations (\ref{eq: incre any one traces of r's}) and
(\ref{eq: scaling Q with F}) simplify considerably at the band center
$\varepsilon=0$. At the band center, the scattering matrix is
Hermitian, and hence $r = r^{\dag}$. Restricting our attention to
single traces and products of two traces, we find the scaling equations
\bleq 
\begin{eqnarray} \label{eq:Rn}
  {\gamma \ell \over \beta} \partial_L 
  \left \langle \mbox{tr}\, r^n \right \rangle &=&
  - {n \gamma \over \beta}
  \left \langle \mbox{tr}\, r^n \right \rangle
  + {n \over 2} \sum_{k=0}^{n} 
  \left \langle \mbox{tr}\, r^{n-k+1}\, \mbox{tr}\, r^{k+1} \right \rangle
  + {n \over 2} \sum_{k=1}^{n-1}
  \Big(
  \left \langle \mbox{tr}\, r^{n-k-1}\, \mbox{tr}\, r^{k-1} \right \rangle
  - 2 \left \langle \mbox{tr}\, r^{n-k}\, \mbox{tr}\, r^{k} \right \rangle
  \Big)
  \nonumber \\ && \mbox{} +
  {n \over 2}\left[{{2-\beta\over\beta}}-{2(1-\eta)\over \beta N}\right]
  \Big[ (n+1)
  \left \langle \mbox{tr}\, r^{n+2} \right \rangle
  + (n-1) \left \langle \mbox{tr}\, r^{n-2} \right \rangle
  - 2 (n-1) \left \langle \mbox{tr}\, r^{n} \right \rangle
  \Big], \\
  {\gamma \ell \over \beta} \partial_L 
  \left \langle \mbox{tr}\, r^m\, \mbox{tr}\, r^n \right \rangle &=&
  {\gamma \ell \over \beta} 
    \Big[ 
    \left\langle
    \mbox{tr}\, (\partial_L \mbox{tr}\, r^m) r^n
    \right\rangle 
    +
    \left\langle
    \mbox{tr}\, r^m (\partial_L \mbox{tr}\, r^n) 
    \right\rangle 
    \Big]
   + {2mn \over \beta} 
   \left\langle \mbox{tr}\, r^{m+n-2}(1-r^2)^2 \right\rangle
   \nonumber \\ && \mbox{} - 
     {2mn(1-\eta) \over \beta N}
     \left\langle \mbox{tr}\, r^{m-1}(1-r^2)\, \mbox{tr}\, r^{n-1}(1-r^2)
     \right\rangle. \label{eq:RnRm}
\end{eqnarray}
\eleq\noindent
Here $(\gamma \ell/\beta) \partial_L \mbox{tr}\, r^n$ is the r.h.s.\ of
Eq.\ (\ref{eq:Rn}) with the omission of the disorder averaging
brackets. (If $n = 1$, the last term in Eq.\ (\ref{eq:Rn}) should be
omitted.) 
[Alternatively, one could have used the Fokker-Planck equation
(\ref{eq:FP ch}) to derive these scaling equations. Both methods agree as we
have verified explicitly.]

The average and variance of the conductance $g=N-{\rm tr}\, r^2$ can be
computed by straightforward solution of Eqs.\ (\ref{eq:Rn}) and
(\ref{eq:RnRm}) using the decoupling scheme of Sec.\ \ref{sec:RG
equations}. The result is, up to corrections of order $N^{-1}$,
\begin{mathletters}
\label{eq: weak-localization for chiral}
\begin{eqnarray}
  \left\langle g\right\rangle &=&
  {N\over s+1}
  + {2-\beta\over\beta} {s^2 \over (1+s)^3},
  \label{eq:weak loc g}
\\
  {\rm var}\, g &=&
  {4\over 15\beta} \left[ 1- {6s+1\over(s+1)^6} \right],
\label{eq:var g} 
\end{eqnarray}
\end{mathletters}
where, as before $s = L/\ell$.
For the derivation of these results we needed the following intermediate
results,
\begin{eqnarray*}
  \langle \mbox{tr}\, r^4 \rangle &=&
  {N s^2(3 s^2 + 8 s + 6) \over 3(1+s)^4}, \\
  \langle \mbox{tr}\, r^6 \rangle &=&
  {
N s^3(
 15 s^4
+ 82 s^3  
+ 177 s^2  
+ 180 s 
+ 75 
) 
\over 15(1 + s)^7},
\end{eqnarray*}
up to corrections of order $N^{0}$ and
\begin{eqnarray*}
  \left\langle 
  {\rm tr}\,r\, {\rm tr}\,r
  \right\rangle &=&
{2\over3\beta}\left[1-{1\over(s+1)^3}\right],
\\ \left\langle \mbox{tr}\, r\, \mbox{tr}\, r^3 \right \rangle &=&
{2\over15\beta}
\left[
4-
{
 5s^3
+15 s^2
+24 s
+4
\over
(s+1)^6}
\right],
\end{eqnarray*}
up to corrections of order $N^{-1}$.

In the diffusive regime $\ell \ll L \ll N \ell$ we observe that the
variance of the conductance at $\varepsilon=0$ is twice the value taken
in the standard case, for $\varepsilon$ far away from $0$, cf.\ Eq.\
(\ref{eq:var g st}). The result that the presence of the extra chiral 
symmetry leads to a doubling of
conductance fluctuations was found previously for the random flux model
(corresponding to our case $\beta=2$) from numerical
simulations\cite{Ohtsuki93} and from an exact solution of the
Fokker-Planck equation (\ref{eq:FP ch}).\cite{Mudry99} The factor
two decrease in the fluctuations as the chiral symmetry is broken, 
is reminiscent of the factor two decrease of the conductance fluctuations
upon breaking time-reversal symmetry\cite{BeenakkerReview} or upon breaking
a spatial symmetry.\cite{BrouwerBeenakker95,BarangerMello96}

According to Eq.\ (\ref{eq:weak loc g}), application of a magnetic
field has an effect on the average conductance, but this effect
vanishes in the diffusive limit $\ell\ll L \ll N\ell$, i.e., $s\gg1$.
{In other words, there is no weak-localization correction
to the conductance in the diffusive regime.} 
It is instructive to note a coincidence between the
$\beta$-dependence of the average conductance $\langle g \rangle$ and 
the $\beta$-dependence of the localization length $\xi$ which was
considered in the previous section. In the case of the random hopping
model at zero energy, there is no weak-localization correction, and the
localization length $\xi$ does not depend on the presence or absence of
time-reversal symmetry. On the other hand, without chiral symmetry (for
large energies), the negative correction to the average conductance for
$\beta=1$ foreshadows the localization transition, which occurs on a
length scale $\xi_{\rm st}$ that is proportional to $\beta$, i.e.,
localization takes place twice as fast without than with a
time-reversal symmetry breaking magnetic field.
Absence of weak-localization correction to the conductance
had been pointed out by Gade and Wegner in their study 
of a (two-dimensional)
non-linear-$\sigma$ model implementing the chiral symmetry.\cite{Gade93} 
(See also Ref.~\onlinecite{Fabrizio00}.)

Finally, notice that there is no precursor in Eqs. 
(\ref{eq: weak-localization for chiral}) 
of the even-odd effect seen in the localized
regime. This agrees with the exact solution for $\beta=2$, where it was
found that the even-odd effect is non-perturbative in the expansion
parameter $L/N\ell$.\cite{Mudry99}

To order $N$, the average conductance is the same as in the case of a
wire without chiral symmetry. Differences show up only to order $N^0$,
where we find that there are no weak-localization corrections for the
chiral case. This is not a coincidence that is limited to the average of the
conductance $g = \mbox{tr}\, t^{\dagger} t$. It extends to the averages
of traces of arbitrary powers of $r$ or $t$. To see this and
in order to allow for a
more detailed comparison to the case where chiral symmetry is absent,
we rephrase the scaling equation (\ref{eq:Rn}) in terms of the transmission 
matrix $t$. In the limit of large $N$, one thus obtains
\bleq 
\begin{eqnarray}
  \left(N + {2 - \beta \over \beta} \right) \ell \partial_L
  \left\langle {\rm tr}\, (t^{\dag}t)^n\right\rangle &=&
  - n \sum_{m=1}^{n  } 
  \left\langle {\rm tr}\, (t^{\dag}t)^{n+1-m} \right\rangle
  \left\langle {\rm tr}\, (t^{\dag}t)^{    m} \right\rangle
  + n\sum_{m=1}^{n-1} 
  \left\langle {\rm tr}\, (t^{\dag}t)^{n  -m} \right\rangle
  \left\langle {\rm tr}\, (t^{\dag}t)^{    m} \right\rangle
  \nonumber \\ && \mbox{}
  - n (2n + 1) {{2-\beta\over\beta}}
  \left\langle {\rm tr}\, (t^{\dag}t)^{n+1  } \right\rangle
  + 2 n^2 {{2-\beta\over\beta}}
  \left\langle {\rm tr}\, (t^{\dag}t)^{n    } \right\rangle
  + {\cal O}(N^0).
\label{eq:RG for ga's}
\end{eqnarray}
\eleq 
\noindent
For a quantum wire without the chiral symmetry, 
the leading ${\cal O}(N)$ contribution to $\langle
{\rm tr}\, (t^{\dag}t)^{n}\rangle$ is precisely the same
as the first line of Eq.~(\ref{eq:RG for ga's}).
The weak-localization correction proportional to ${{2-\beta\over\beta}}$
differs in the standard case from the second line of 
Eq.~(\ref{eq:RG for ga's})
as it reads $-n^2 {{2-\beta\over\beta}} 
\langle {\rm tr}\, (t^{\dag}t)^{n}\rangle + n(n-1) {{2-\beta\over\beta}}
\langle {\rm tr}\, (t^{\dag}t)^{n+1}\rangle$.\cite{Mello91} 
Hence, whereas the
solution of Eq.\ (\ref{eq:RG for ga's}) is the same as for an ordinary
quantum wire to leading order in $N$,
\begin{equation}
    \left\langle {\rm tr}\, (t^{\dag}t)^{n    } \right\rangle
 = {N \over 2 s} B(n,1/2),
\end{equation}
where $B(x,y) = \Gamma(x) \Gamma(y)/\Gamma(x+y)$ is the beta
function,\cite{Dorokhov84,MelloPichard89} 
the combination $(2n+1) \langle {\rm tr}\, (t^{\dag}t)^{n+1 }
\rangle - 2 n \langle {\rm tr}\, (t^{\dag}t)^{n    } \rangle$
on the second line of Eq.~(\ref{eq:RG for ga's})
conspires with the coefficient $B(n,1/2)$ to insure the disappearance
of a weak-localization correction for all averages $\langle {\rm tr}\,
(t^{\dag}t)^{n} \rangle$ in the presence of the chiral symmetry. As a
corollary, we find that the average density of the transmission
eigenvalues $x_j$
\begin{equation}
  \rho(x) = \left\langle \sum_{j=1}^{N} \delta(x-x_j) \right\rangle
\end{equation}
has no weak-localization correction as well.

With these results, it is little work to compute the average shot
noise power
$\langle p \rangle =
 \langle \mbox{tr}\, t^{\dagger} t(1-t^{\dagger} t)\rangle$
and its weak-localization correction,
\begin{eqnarray}
  \left\langle p\right\rangle &=&
  {N\over 3} \left[ {1\over s+1}- {1\over(s+1)^4} \right]
  \nonumber \\ && \mbox{}
  + {2-\beta\over\beta}
  \left[ {s^2 \over 3 (1+s)^3} - {7 s^2 \over 3 (1+s)^6} \right].
  \label{eq:weak loc P}
\end{eqnarray}
Just like in the case of the conductance there is no weak-localization
correction in the diffusive regime $\ell \ll L \ll N\ell$.

\subsection{Crossover between the chiral and standard universality classes}
\label{sec:Crossover between the chiral and standard universality classes}

For any nonzero energy $\varepsilon$, the chiral symmetry 
of Eq. (\ref{eq:Schiral}) is broken.
Hence one expects that for a sufficiently long length $L$ of the
quantum wire, its transmission properties will flow to those of the
standard symmetry class. This flow is governed by a crossover length
scale $\ell_{\varepsilon}$ so that for $L \ll \ell_{\varepsilon}$, the
transmission properties are still alike those in the chiral symmetry
class, while for $L \gg \ell_{\varepsilon}$ they resemble those of the
standard symmetry class. 
We distinguish three possible regimes where
this crossover can take place:
\begin{itemize}
\item  The crossover takes place in the ballistic regime,
       $\ell_{\varepsilon} \ll \ell$,
\item  The crossover takes place in the diffusive regime,
       $\ell \ll \ell_{\varepsilon} \ll N \ell$, or
\item  The crossover takes place in the localized regime,
       $\ell_{\varepsilon} \gg N \ell$. This regime cannot be
       treated with the methods used in the paper. For the
       case $N=1$ of a single-channel quantum wire, this
       regime has been studied in
       Refs.\ \onlinecite{Theo76,Eggarter78,Stone81,Balents97,Mathur97}.
\end{itemize}
The full set of scaling equations (\ref{eq: incre any one traces of r's})
and (\ref{eq: scaling Q with F}) can be used to describe the first two
regimes (and the intermediate region between them).
Although the solution of the scaling
equations is straightforward --- within the large-$N$ decoupling
scheme, the scaling equations are linear ordinary differential equations
that can be solved one-by-one (see appendix \ref{sec:The appendix})
--- it is a quite cumbersome task, and
many expressions get quite lengthy. [The expression (\ref{eq: cross tr
rr leading}) for $\langle \mbox{tr}\, r^2 \rangle$ is the only example
whose solution can be represented by a one-line equation.] To simplify
our presentation and to save the reader from those lengthy expressions,
we focus on the regime $\ell \ll \ell_{\varepsilon} \ll N \ell$, where
the crossover takes place inside the regime of diffusive dynamics.

The length scale for the crossover can be identified from Eq.\
(\ref{eq: cross tr rr leading}) as
\begin{equation}
  \ell_{\varepsilon} = \sqrt{{\ell \over 2 \varepsilon}}. 
  \label{eq:ell_e}
\end{equation}
(Here we have neglected $\varepsilon \ell$ with respect to ${i}$ in 
$\sqrt{{i} + \varepsilon \ell}$. 
This is consistent with our 
focus on the regime $\ell_{\varepsilon} \gg \ell$.) 
We note that $\varepsilon$ is none
but the Thouless energy for a diffusive process with diffusion constant
$v_f\ell$ (having momentarily reinstated the Fermi velocity $v_f$) in
a system of linear size $\ell_{\varepsilon}$.
Using the hierarchy of length scales $\ell \ll \ell_{\varepsilon}$ we
then find that the solution of the scaling equations takes a relatively
simple form. For the average and variance of the conductance $g$
we find up to ${\cal O}(N^0)$
\bleq
\begin{eqnarray}
 \left\langle g \right\rangle &=&
   {N \ell \over \sigma \ell_{\varepsilon}}
  -{2-\beta\over\beta}
  \left[ {1\over 3} 
  - {z\coth(z^*{\sigma})  +z^*\coth(z{\sigma})\over 4{\sigma}} \right].
\label{eq:cross mean g}
  \\
  \mbox{var}\, g &=&
  {2\over 15\beta} 
  + {2\over\beta} \left[
  {3z \sigma \coth(z{\sigma}) - 2\over 16{\sigma}^4}
  + {{i}\over 8{\sigma}^2\sinh^2(z^*{\sigma})} 
  + {\rm c.c.}
  \right]. 
\label{eq: cross var g}
\end{eqnarray}
Here we defined $z = 1 + {i}$ and $\sigma = L/\ell_{\varepsilon}$.
For the average of the shot-noise power we find up to ${\cal O}(N^{0})$
\begin{eqnarray}
  \langle p \rangle=
  {N \ell \over 3 \sigma \ell_{\varepsilon}} 
  - {2-\beta\over\beta}
  \left\{ {1\over 45} 
  + \left[ {(3 z-2z^* \sigma^2) \coth(z{\sigma}) \over 24{\sigma^3}}
  + {{i}\over 4{\sigma}^2\sinh^2(z^*{\sigma})} 
  + {\rm c.c.} \right]
  \right\}. \label{eq: cross shot noise}
\end{eqnarray}
For the derivation of these results, we needed the following intermediate
results, all up to corrections of order $N^0$,
\begin{eqnarray}
  \langle{\rm tr}\,(r^{\dag}r)^3\rangle &=& 
  N - {N \ell \over \ell_{\varepsilon}}\,
  {23 \over 15 \sigma},
  \label{eq: R101010}
  \nonumber \\
  \langle{\rm tr}\, r^2 \rangle &=&
  N 
  -{N \ell \over \ell_{\varepsilon}} z^*\coth(z^*{\sigma}),
  \label{eq: R00}
  \nonumber \\
  \langle{\rm tr}\,r^{\dag}r^3\rangle &=& 
  N
  - {N \ell \over \ell_{\varepsilon}}
  \left[
  { (4{\sigma}^2 + {i}) z^*\coth(z^*{\sigma})
  \over   4{\sigma}^2 }
  - {1 \over 2{\sigma}\sinh^2(z^*{\sigma})}
  \right],
  \label{eq: R1000}
  \nonumber \\
  \langle{\rm tr}\,r^{\dag 2}r^2\rangle &=& 
  N - {N \ell \over \ell_{\varepsilon}}
  \left[
  {4 \sigma - z^*\coth(z{\sigma}) - z\coth(z^*{\sigma}) 
  \over 2{\sigma}^2} \right],
  \label{eq: R1100}
  \nonumber \\
  \langle{\rm tr}\,(r^{\dag}r^2)^2\rangle &=&
  N - {N \ell \over \ell_{\varepsilon}}
  \left[
  {(16 \sigma^4 z^{*} + 8 \sigma^2 z + z^{*}) \coth(z^{*} \sigma)
  \over 16 \sigma^4}
  - {z\coth(z{\sigma})\over 4{\sigma}^4} 
  - {8 \sigma^2 - 2 z \sigma \coth(z^{*} \sigma) + 5 {i}
  \over 8 \sigma^3 \sinh^2 (z^* \sigma)} \right],
  \label{eq: R100100}
  \nonumber \\
  \langle{\rm tr}\,(r^{\dag}r)^2r^2\rangle &=&
  N - {N \ell \over \ell_{\varepsilon}}
  \left[
  {(96 \sigma^4 z^{*} + 40 z \sigma^2 + 3 z^*) \coth (z^* \sigma)
  \over 96 \sigma^4}
  - {40 \sigma^2 - 6 z \sigma \coth(z^{*} \sigma) + 3 {i} \over
  48 \sigma^3 \sinh^2(z^* \sigma)} \right].
  \label{eq: R101000}
\end{eqnarray}
\eleq 

In the limit $\sigma \to 0$, corresponding to $L \ll
\ell_{\varepsilon}$, the weak-localization corrections to the
conductance and the shot noise power and the conductance fluctuations
approach their values for the chiral symmetry class
cf.\ Eqs.\ (\ref{eq: weak-localization for chiral}) and (\ref{eq:weak
loc P}). For $\sigma \gg 1$, corresponding to $L \gg
\ell_{\varepsilon}$, one verifies that the values corresponding to
the standard symmetry class are recovered.

In this subsection, we have
described the effect of a finite energy by a crossover length scale
$\ell_{\varepsilon}$. For the conductance and the shot
noise power the limits of large and small $\varepsilon$ correspond to
the limits of $L$ large or small compared to $\ell_{\varepsilon}$.
However, upon inspection
of Eq.\ (\ref{eq: cross tr rr leading}) or (\ref{eq: R101000}) one
observes that traces like $\mbox{tr}\, r^2$ that contain different
numbers of $r$'s and $r^{\dag}$'s, do not approach their large-energy
limits $\langle \mbox{tr}\, r^2 \rangle = 0$ as
$L \to \infty$.\cite{footstrongdisorder} 
The origin of this difference is
that the reflection matrix is dominated by (interference of) 
paths that only enter a
distance of the order of a mean free path $\ell$ into the quantum wire,
while the conductance and the shot noise power 
depend on quantum interference throughout the entire
wire. Hence, as long as $\ell_{\varepsilon} \gg \ell$, the finite
energy cannot alter the interference of most paths that contribute to
$r$. Hence, to judge whether the finite energy is relevant for the
traces of reflection matrices, one has to compare
$\ell_{\varepsilon}$ to $\ell$ instead of $L$.

We are now ready to define what is meant by ``$\varepsilon$
sufficiently large'' in the crossover from the chiral symmetry class to
the standard symmetry class. As far as quantum interference corrections
to the transmission properties are concerned, the results of this
subsection show that ``$\varepsilon$ sufficiently large'' corresponds
to the inequality of length scales $L \gg \ell_{\varepsilon}$, or
equivalently, $\varepsilon \gg \ell/L^2$.  However for reflection
traces like $\mbox{tr}\, r^2$, a much more strict criteria is needed,
$\ell_{\varepsilon} \ll \ell$, or $\varepsilon \ll \ell$. In the next
section, these criteria, as well as the functional forms 
(\ref{eq:cross mean g}), (\ref{eq: cross var g}) 
for the crossover will be compared to numerical simulations.

\section{Numerical simulations}
\label{sec:Numerical simulations}

In this section we report on numerical simulations of the conductance
of a quantum wire with random hopping only, and compare them 
with the theory of sections
\ref{sec:Microscopic model and scattering matrix}-\ref{sec:Diffusive
regime}.

The simulations are for the random hopping model on a square lattice,
described by the Schr\"odinger equation, 
\begin{eqnarray}
\varepsilon\psi_{m,j}&=&
-t_{m,j-1;\perp}\psi_{m,j-1}-t_{m,j;\perp}^*\psi_{m,j+1}
\nonumber\\&&
-t_{m-1,j;\parallel}\psi_{m-1,j}-t_{m,j;\parallel}^*\psi_{m+1,j},
\label{Sch num model}
\end{eqnarray}
where $\psi_{m,j}$ is the wave function at the lattice site $(m,j)$.
A site is labeled by the chain index $j=1,\ldots,N$ 
and by the column index $m$.
We impose open boundary conditions in the transverse direction,
$t_{m,0;\perp}=t_{m,N;\perp}=0$. The system consists of a disordered
region ($0 < m < L$), coupled to the left and right to perfect
leads ($m<1$ and $m>L$).
In the leads, the longitudinal and transverse hopping amplitudes are
$t_{m,j;\parallel}=1$ and $t_{m,j;\perp}=t$, where $0<t\le1$.
With this choice, there is a window of energies
$-1+t < \varepsilon <1-t$ around the band center, where the number of 
transmission channels does not depend on energy 
(and equals the number of chains $N$). 
In the disordered region, the hopping amplitudes are taken from 
a distribution centered around the values $t_{m,j;\parallel}=1$ 
and $t_{m,j;\perp}=t$ for the leads.
We consider two types of randomness, that we refer to as the
real random hopping and random flux models. 
\begin{itemize}

\item In the real random hopping
(RRH) model, the hopping amplitudes $t_{m,j;\perp}$ and
$t_{m,j;\parallel}$
are chosen uniformly and independently in the intervals
$-t(1-\delta) < t_{m,j;\perp} < t(1+\delta)$ 
and
$1 - \delta < t_{m,j;\parallel} < 1 + \delta$, 
respectively, where $\delta$ measures the disorder strength.
A uniform magnetic field with a flux $\phi_{\rm pl}$ through each
plaquette is modeled by multiplication of $t_{m,j;\parallel}$ with a
Peierls phase $e^{2 \pi {i} \phi_{\rm pl} (j-1)}$. 

\item In the random flux (RF) model, the longitudinal hopping amplitudes 
$t_{m,j;\parallel} = 1$, while the transverse hopping amplitudes
$t_{m,j;\perp}$ are complex numbers 
$t_{m,j;\perp} = t e^{{i}\theta_{m,j}}$.
Here the $\theta_{m,j}$ are chosen such that the fluxes 
$\phi_{m,j}= \theta_{m,j} - \theta_{m-1,j}$ 
are independently and uniformly distributed in the interval 
$-\pi p < \phi_{m,j} < \pi p$, 
where $p$ is a measure for the strength of the disorder. 
\end{itemize}
In the random
flux model, the parameter $\eta=0$, see Ref.\ \onlinecite{Brouwer99NON};
in the real random hopping model the precise value of $\eta$ is 
not known. However, nonzero $\eta$ (of order $N^0$ by assumption)
will only give
rise to corrections of relative order $1/N$, ($1/N^2$ for the 
average conductance), which can be neglected for large $N$. 
Since the statistics of the conductance in the RF model in a
quasi-one-dimensional geometry has been studied extensively
in Ref.\ \onlinecite{Mudry99} at and away from the band center
$\varepsilon=0$, we restrict our attention here to the crossover
as a function of energy. 

The wavefunctions that solve the Schr\"odinger equation 
(\ref{Sch num model}) at energy $\varepsilon$ can be written as
\begin{eqnarray}
\psi_{m,j} &=&
\sum^N_{\nu=1}\frac{1}{\sin k_\nu} \left[
e^{{i}k_\nu m}\sin(q_\nu j) \psi_\varepsilon^{\rm iL}(\nu)
  \right. \nonumber \\ && \left. \mbox{}
+e^{-{i}k_\nu m}\sin(q_{N+1-\nu} j) \psi_\varepsilon^{\rm oL}(\nu)\right]
\label{psi^L(nu)}
\end{eqnarray}
in the left lead and as
\begin{eqnarray}
\psi_{m,j} &=&
\sum^N_{\nu=1}\frac{1}{\sin k_\nu} \left[
e^{-{i}k_\nu m}\sin(q_{N+1-\nu} j) \psi_\varepsilon^{\rm iR}(\nu)
  \right. \nonumber \\ && \left. \mbox{}
+e^{{i}k_\nu m}\sin(q_\nu j) \psi_\varepsilon^{\rm oR}(\nu)\right]
\label{psi^R(nu)}
\end{eqnarray}
in the right lead,
where the wave number $k_\nu>0$ is determined from
$\varepsilon=-2\cos k_\nu-2t\cos q_\nu$ with $q_\nu=\pi\nu/(N+1)$.
With this parameterization, the 
definition of the scattering matrix $S_{\varepsilon}$
and its symmetries are the
same as in Sec.\ \ref{sec:Microscopic model and scattering matrix}.

For each realization of the disorder, the dimensionless conductance
$g$ is computed from the Landauer formula (\ref{eq:Landauer}).
The recursive Green's function\cite{MacKinnon83,Ando89,Baranger91} 
method is used to calculate $S_{\varepsilon}$. (Application of the
method to the random hopping or random flux models is discussed
in Ref.\ \onlinecite{Mudry99}.)
Our numerical simulations use the
parameters $t = 0.6$, $\delta = 0.2$, and $p = 0.3$, for the
RRH and RF models.

\subsection{Localized regime in the RRH model}
\label{subsec:Localized regime in the RRH model}

\begin{figure}
\centerline{\epsfxsize=85mm\epsfysize=69mm\epsfbox{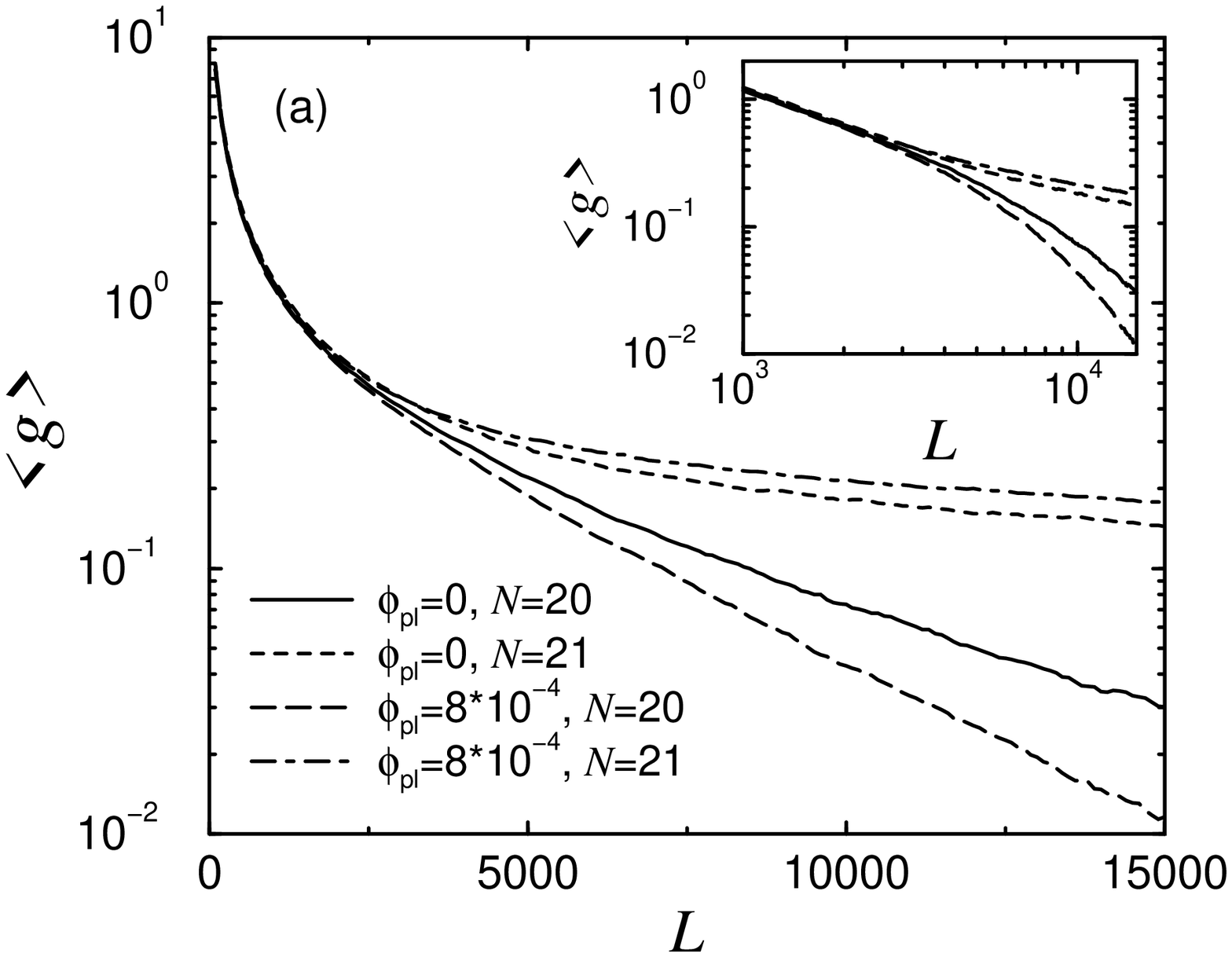}} 
\vspace{-0.5cm}
\centerline{\epsfxsize=85mm\epsfysize=69mm\epsfbox{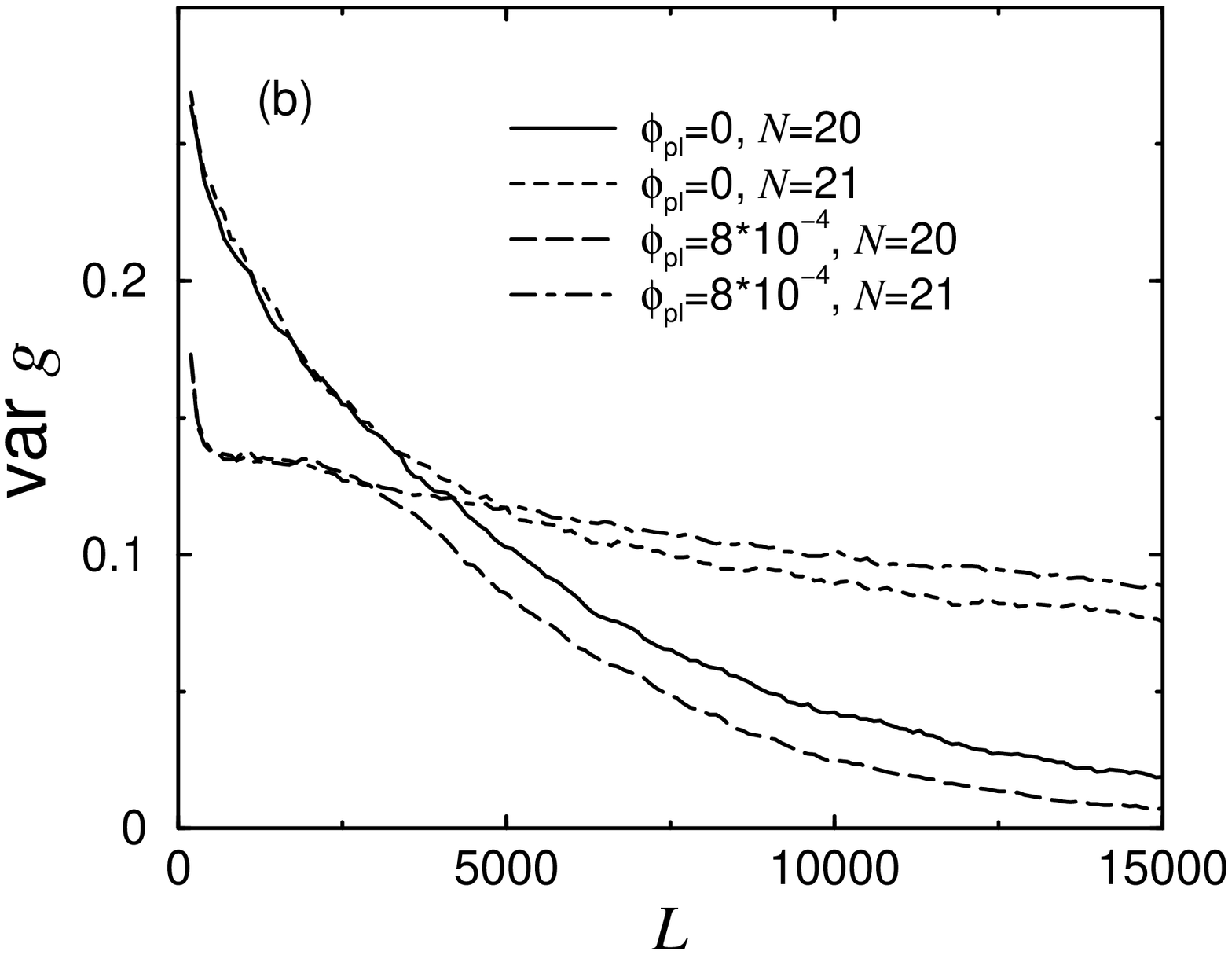}} 
\caption{
Mean (a) and variance (b) of the conductance 
for $N=20$ and 21 at the band center
$\varepsilon=0$ with and without magnetic field in the RRH model.
Averaging over $2\times 10^4$ realizations of disorder is performed.
}
\label{fig:even-odd of g}
\end{figure}

In the localized regime, the even-odd effect
manifests itself most dramatically. Taking an average over
$2 \times 10^4$ realizations of the disorder, we
have computed the mean and variance of $g$ at the band center
$\varepsilon=0$ for the RRH model with $N=20$ and $21$, and with
and without a time-reversal breaking magnetic field, see Fig.\
\ref{fig:even-odd of g}. 
The magnetic field corresponds to a flux $\phi_{\rm pl} =
8 \times 10^{-4}$ per plaquette, or $\sim 1$ flux
quantum per $50$ lattice spacings along the chain, so that 
time-reversal symmetry is broken for all but the shortest 
wire lengths shown in Fig.\ \ref{fig:even-odd of g}.
For odd $N(=21)$ both $\langle g\rangle$ and ${\rm var}\,g$ decrease
algebraically whereas they decay exponentially for even $N(=20)$.
We observe that, for odd (even) $N$ and fixed $L$,
$\langle g\rangle$ and $\mbox{var}\, g$ are larger (smaller) 
in the presence of a magnetic field, $\beta=2$, than without, $\beta=1$,
in agreement with 
Eqs.~(\ref{eq:chiral g}) and (\ref{eq:chiral g odd}).
Note that for small $L$, $\mbox{var}\, g$ is 
$L$-independent for the chiral unitary class, while $\mbox{var}\, g$
decreases linearly with $L$ for small $L$ in the chiral orthogonal 
class. Similar $L$-dependencies for small $L$ have been obtained
for the standard symmetry classes, see Ref.\ \onlinecite{Mirlin94}.

\begin{figure}
\epsfxsize=85mm
\epsffile{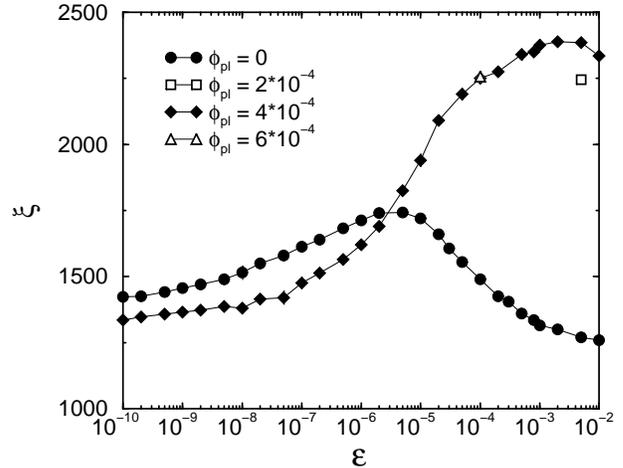}
\caption{
Localization length $\xi$ as functions of energy
$\varepsilon$ for $\phi_{\rm pl}=0$, $2\times10^{-4}$,
$4\times10^{-4}$, and $6\times10^{-4}$
in the RRH model with $N=20$ chains.
Averaging over $500$ realizations of disorder is performed.
}
\label{fig:loclength}
\end{figure}

\begin{figure}
\epsfxsize=85mm
\epsfysize=69mm
\epsffile{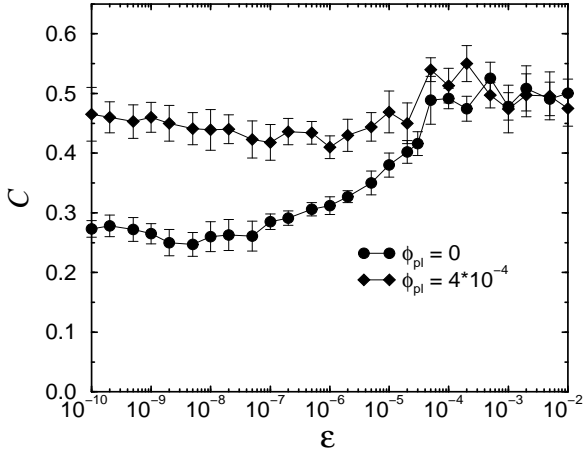}
\caption{
Ratio $C = - \lim_{L \to \infty} \langle\ln g\rangle/ {\rm var}\ln g$ 
versus $\varepsilon$ for $\phi_{\rm pl}=0$ and $4\times10^{-4}$
in the RRH model with $N=20$ chains and $L=4\times10^5$.
The disorder average is taken over $500$ realizations.
}
\label{fig:ratio}
\end{figure}

Results for the crossover from the chiral universality
classes to the standard ones as a function of energy are shown in
Figs.\ \ref{fig:loclength} and \ref{fig:ratio}. Figure
\ref{fig:loclength} shows the energy dependence of the localization 
length $\xi = -2 \lim_{L \to \infty} L/\langle \ln g \rangle$ [cf.\
Eq.\ (\ref{eq:chiral_even_mean lng})] for $N=20$; Fig.\
\ref{fig:ratio} shows numerical data for the
ratio $C=- \lim_{L \to \infty} \langle \ln g\rangle/{\rm var}\, 
\ln g$.  Here, the averages were taken over $500$--$1000$ realizations
of the disorder and magnetic fields corresponding to fluxes
$\phi_{\rm pl}=2,4,6\times 10^{-4}$ per plaquette, respectively, 
have been used.

In the absence of a magnetic field,
$\xi(\varepsilon)$ shows non-monotonic behavior with a maximum around 
$\varepsilon\approx5\times10^{-6}$, while, within $10\%$, the
localization length $\xi$ is
the same in the chiral orthogonal class ($\varepsilon = 0$) and in
the standard orthogonal class ($\varepsilon \gtrsim 10^{-4}$ for the
choice of parameters in the simulations), in agreement
with Sec.\ \ref{sec:localized regime}. As we discussed
in Sec.\ \ref{sec:localized regime}, the fact that 
$\xi(\varepsilon=0) = \xi(\varepsilon \gg 0)$ in  the absence of a 
magnetic field, could be interpreted as the result of a cancellation 
of two effects: 
The presence of an extra symmetry at the band center
(the chiral symmetry), which
tends to make $\xi$ shorter than away from the band center, and
the enhancement of the DoS at the band center,
which tends to make $\xi$ larger.\cite{Brouwer99DOS} 
Apparently, these two competing effects are not balanced in the
crossover region, thereby giving rise to the non-monotonic energy
dependence to $\xi$ displayed in Fig.\ \ref{fig:loclength}.
Such a non-monotonicity of $\xi(\varepsilon)$ is reminiscent
of the non-monotonic voltage dependence of differential 
conductance found in a normal-metal/superconductor
microbridge.\cite{NazarovStoof}
In the presence of a magnetic field, $\xi$ increases by a factor
$\sim 1.8$ in the crossover from the chiral-unitary to the standard
unitary symmetry class, which is slightly less than the factor $2$
predicted in Eq.\ (\ref{eq:loc length standard}). Note that the
increase in localization length is most rapid
around the same energy scale $\varepsilon_{\rm c}\approx5\times10^{-6}$
for which $\xi$ reaches its maximum, $\xi_{\rm c}$,
in the absence of a magnetic field. Moreover,
$\varepsilon_{\rm c}$ is related to $\xi_{\rm c}$ by Thouless relation
$\xi_{\rm c}\sim\sqrt{\ell/\varepsilon_{\rm c}}$ where
the mean free path $\ell$ is obtained by dividing the localization 
length $\xi(\varepsilon=10^{-10})$ 
by $N=20$ in Fig.\ \ref{fig:loclength}.

A plot of the ratio $C = - \lim_{L \to \infty} \langle\ln g\rangle /
{\rm var}\ln g$ is shown in Fig.\ \ref{fig:ratio}. In the standard
symmetry classes, $C$ takes the universal value $C = 1/2$, while
in the chiral classes one has\cite{Brouwer99NON}
\begin{eqnarray}
  C &=&\frac{\beta N/2}{N+\left(1-\frac{2}{\pi}\right)(N-2+2\eta)}
  \nonumber \\ &=& {\beta \over 4 - 4/\pi} + {\cal O}(1/N).
\label{ratioC}
\end{eqnarray}
The data shown in Fig.\ \ref{fig:ratio} confirm that $C = 0.5$ in the
standard symmetry classes (corresponding to $\varepsilon \gtrsim
10^{-4}$ for the parameters of our simulation). However, for the
chiral symmetry classes, a $20\%$ discrepancy with Eq.\ 
(\ref{ratioC}) is found. 

While the numerical simulations for the localized regime qualitatively
confirm the theory of Sec.\ \ref{sec:localized regime}, 
quantitative agreement is only up to
$\sim 20 \%$. As a possible source of this discrepancy, we point to
the fact that the simulations are done for an appreciable disorder
strength $\delta = 0.2$, while the theory is derived for weak
disorder, corresponding to $\delta \to 0$. Hence, the system
cannot be considered truly (quasi) one-dimensional, and corrections
from two-dimensional dynamics on shorter length scales need to be
taken into account. Another cause of the observed discrepancies
could be the uncertainty of the precise value of $\eta$ for the RRH
model. While we believe that $\eta$ should not affect the
conductance distribution significantly for large $N$, it remains
difficult to make a quantitative assessment of finite-$N$ corrections
as long as $\eta$ is unknown. At this moment, we are not
aware of a direct way to obtain $\eta$ from the numerical
simulations. 

\subsection{Diffusive regime in the RRH and RF models}

\begin{figure}
\epsfxsize=85mm
\epsfysize=69mm
\epsffile{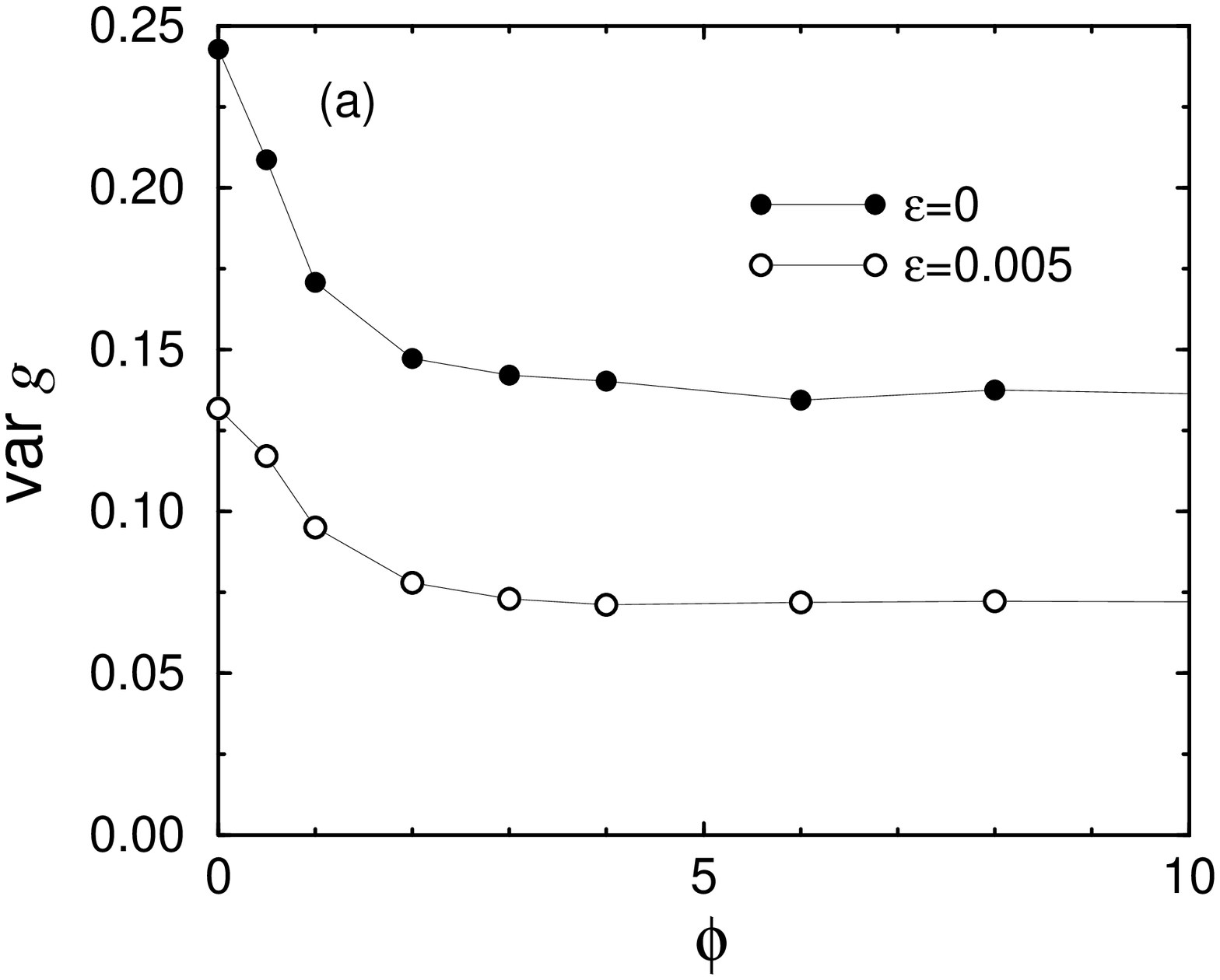}
\vspace{-0.5cm}
\epsfxsize=85mm
\epsfysize=69mm
\epsffile{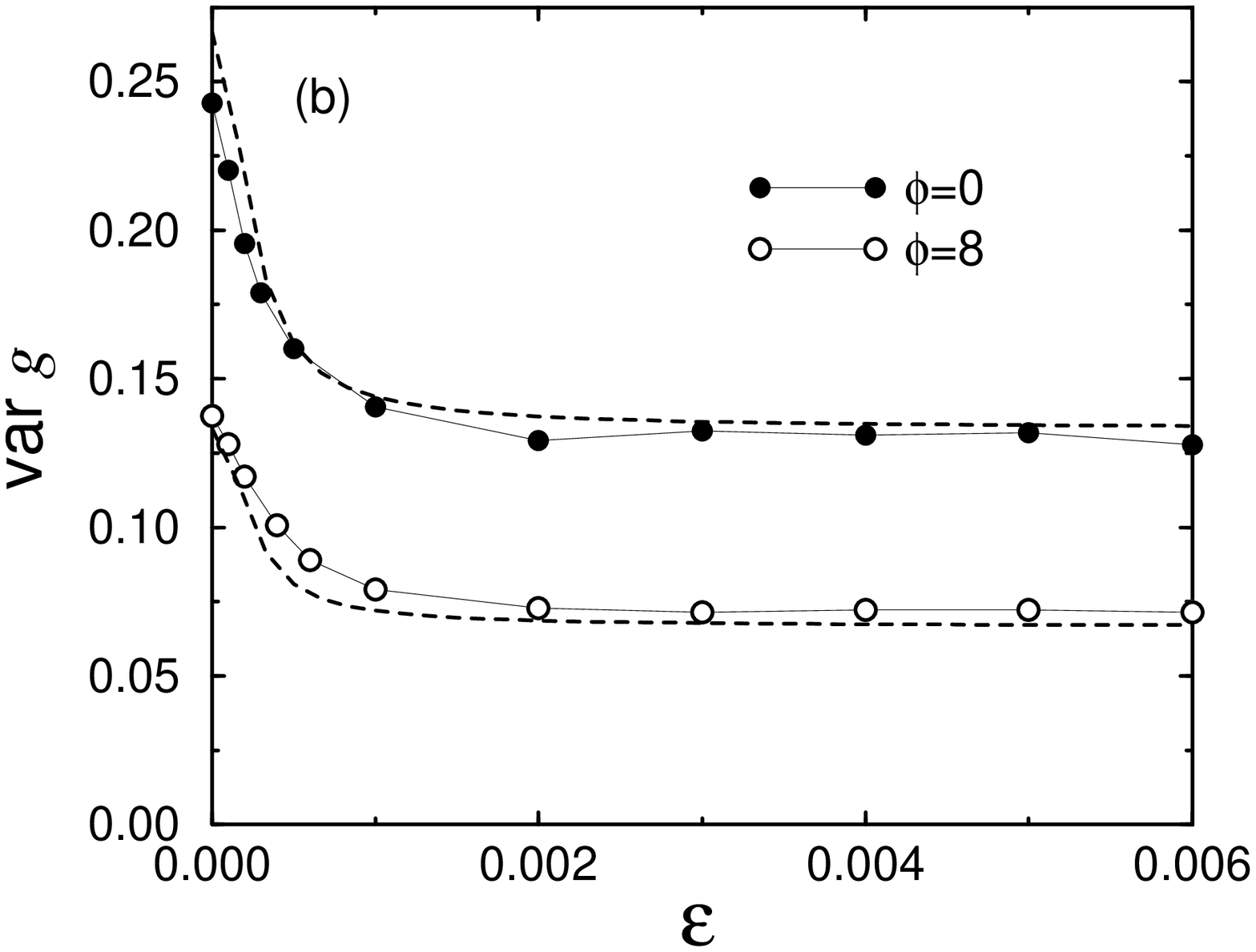}
\caption{ \label{fig:varg_e}
Variance of $g$ for the RRH model
versus the total flux $\phi$ through the disordered
wire (a) and versus energy $\varepsilon$ (b). In (a), $\mbox{var}\, g$
is shown for two values of the energy, $\varepsilon=0$ and 
$\varepsilon=0.005$, corresponding to the presence and absence
of chiral symmetry; in (b), $\mbox{var}\, g$ is shown for two
values of the flux $\phi = 0$ and $\phi = 8$, 
corresponding to the presence and absence of time-reversal symmetry. 
The dashed line in (b) is the theoretical curve 
(\protect\ref{eq: cross var g}),
with $\sigma^2 = 1.5 \times 10^4 \varepsilon$. The
simulations are performed for $N=45$, $L=800$, and an averaging over 
$10^4$ realizations of the disorder is performed.
}
\label{fig:varg_phi}
\end{figure}

\begin{figure}
\epsfxsize=85mm
\epsfysize=69mm
\epsffile{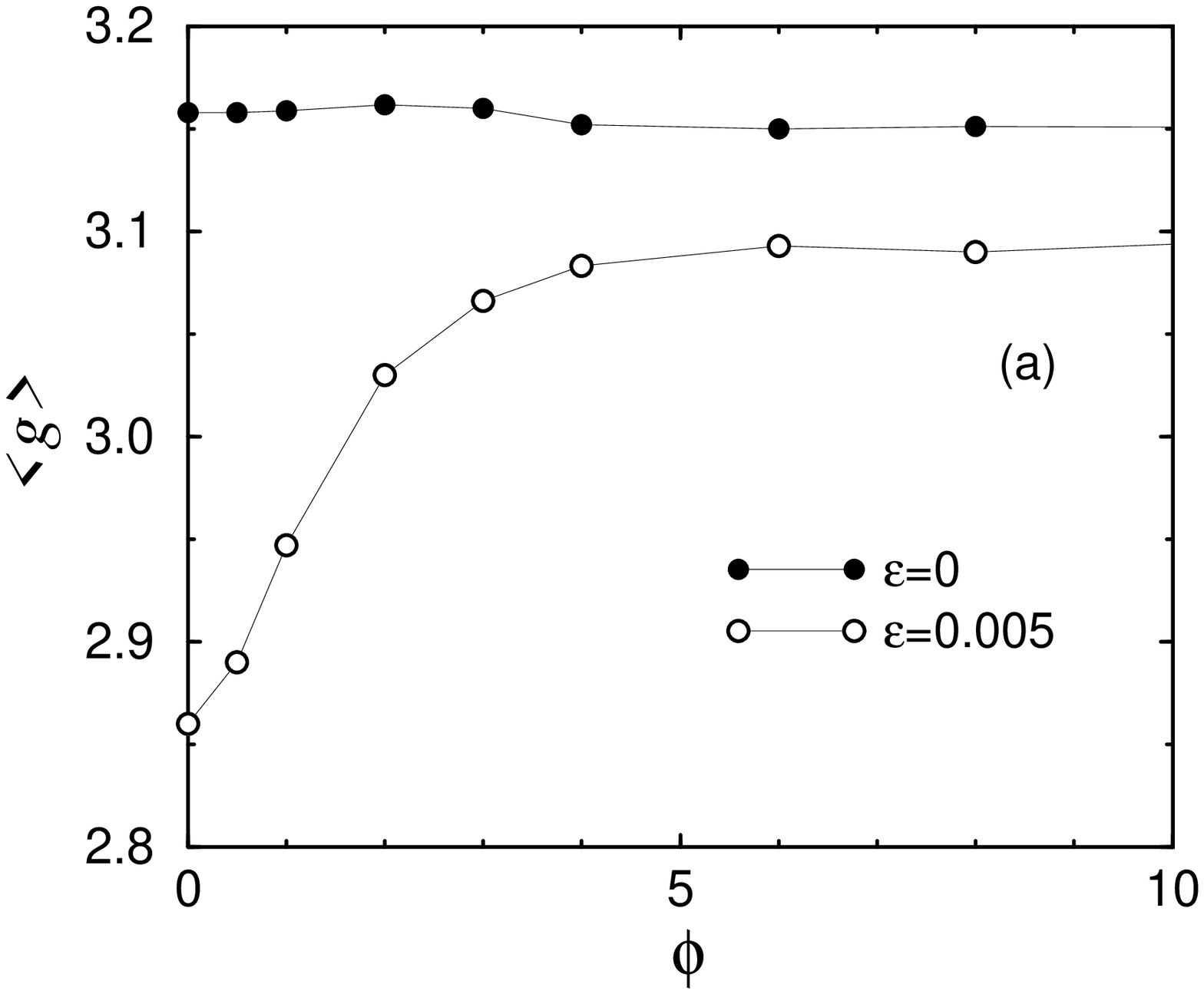}
\vspace{-0.5cm}

\epsfxsize=85mm
\epsfysize=69mm
\epsffile{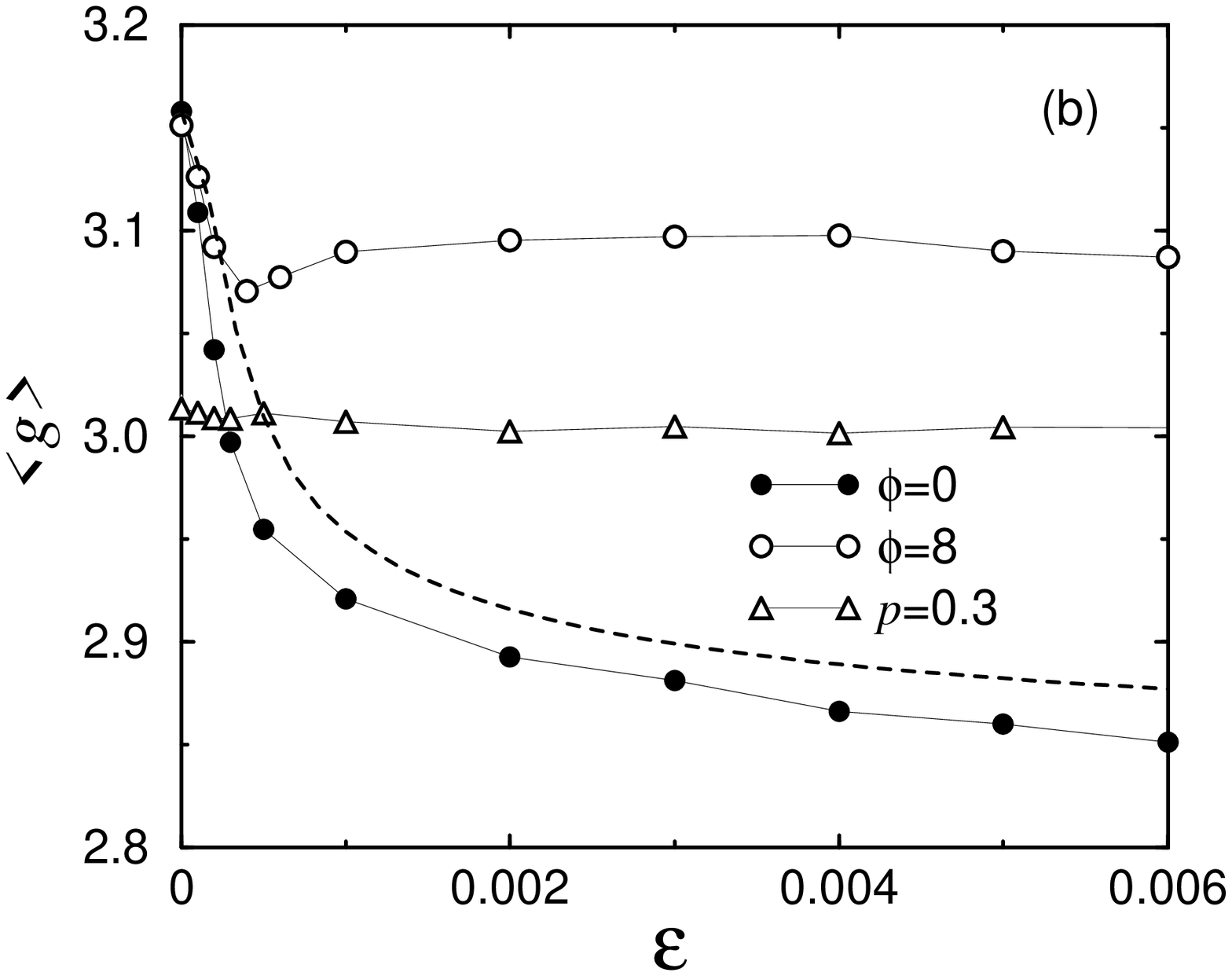}
\caption{
Mean conductance as a function of the total flux $\phi$ (a) and
of the energy $\varepsilon$. In (a), data are shown for
$\varepsilon=0$
and $\varepsilon=0.005$. In (b), $\langle g
\rangle$ is shown for $\phi=0$ and $\phi=8$ for the RRH model
(circles) and for the RF model at $p=0.3$ (triangles). All
simulations were done for $N=45$ and $L=800$. The average was
performed over $10^4$ realizations of the disorder.
The dashed line in (b) is taken from Eq.\ 
(\protect\ref{eq:cross mean g}) with $\beta=1$ and 
$\sigma^2=1.5\times10^4\varepsilon$.
}
\label{fig:g_e}
\end{figure}

We next consider the crossover from the chiral symmetry classes to
the standard symmetry classes in the diffusive regime $\ell\ll
L\ll \xi$. In Figs.\ \ref{fig:varg_phi} and \ref{fig:g_e} we show
numerical simulations of the average and variance of the
conductance as a function of the energy $\varepsilon$ and the
magnetic flux $\phi = L(N-1) \phi_{\rm pl}$ through the disordered 
part of the wire. 
The simulations are performed with $N=45$ and $L = 800$, in order
to ensure that the conditions $\ell\ll L\ll\xi$ for diffusive
transport and $N \ll L$ for quasi-one-dimensionality 
are both met. The ensemble average is taken over $10^4$ samples.

Numerical results for the variance of the conductance in the
RRH model versus $\phi$ and $\varepsilon$ are shown
in Fig.\ \ref{fig:varg_phi}. 
The numerical data of $\mbox{var}\, g$ versus $\phi$ (Fig.\
\ref{fig:varg_phi}a) agree within
$10\%$ with the theoretical predictions $\mbox{var}\, g = 4/15$
($2/15$) for $\phi = 0$ and $\mbox{var}\, g = 2/15$ ($1/15$) 
for $\phi \gg 1$ for $\varepsilon=0$ ($\varepsilon \gg 0$). The
crossover between the orthogonal and unitary classes happens for
$\phi \sim 1$, both with chiral symmetry ($\varepsilon = 0$),
and without ($\varepsilon = 0.005$). The $\varepsilon$-dependence 
of $\mbox{var}\, g$ is shown in 
Fig.\ \ref{fig:varg_e}b, together with the theoretical result
(\ref{eq: cross var g}), where we fitted the crossover energy
scale that enters into the definition of $\sigma =
L/\ell_{\varepsilon}$, cf.\ Eq.\ (\ref{eq:ell_e}). Again we find
quantitative agreement well within $10\%$. (The fact that the
numerical data for $\phi=0$ are below the theoretical curve
can probably be attributed to the suppression of $\mbox{var}\, g$
as $L$ approaches the localization length $\xi$, see the remark in
the discussion of Fig.\ \ref{fig:even-odd of g}.)

Numerical results for the average conductance are shown in Fig.\
\ref{fig:g_e}. All data shown are for the same length $L=800$ and
for the same number of channels $N=45$. For weak disorder one
can ignore the $\phi$ and $\varepsilon$ dependence of the mean
free path $\ell$, and hence of the Drude term in the conductance.
The only effect of a 
variation of $\varepsilon$ or $\phi$ is thus to change the symmetry of 
the quantum wire, which affects the weak-localization
correction to the conductance. According to Eq.\ (\ref{eq:cross mean
g}), we expect a nonzero
weak-localization correction $\delta g$ in the standard orthogonal 
symmetry class, i.e., for $\phi=0$ and $\varepsilon \neq 0$, while
$\delta g = 0$ if time-reversal symmetry is broken ($\phi \gtrsim 1$)
or if the chiral symmetry is present ($\varepsilon=0$). 
This behavior
is confirmed in Fig.\ \ref{fig:g_e}. However, quantitatively, the
numerical results differ $\sim 30\%$ from 
Eq.\ (\ref{eq:cross mean g}). In addition, 
Fig.\ \ref{fig:g_e} shows a small $\varepsilon$ dependence 
of $\langle g \rangle$ at large magnetic fields that cannot be
accounted for within the theory of Sec.\ \ref{sec:Diffusive regime}. 
In particular, note the cusp-like structure
at small $\varepsilon$ in the $\phi=8$ data in Fig.\ \ref{fig:g_e}b.
This effect seems to be too large to
be explained by a spurious $\varepsilon$-dependence of the mean 
free path $\ell$. Since the feature at small $\varepsilon$ is
suppressed at larger lengths $L$, a possible cause might be a contact
resistance effect. (Contact resistance is known to play a role for
disordered normal-metal--superconductor junctions, when the
particle-hole degeneracy is destroyed by a finite voltage or by
a magnetic field.\cite{BrouwerBeenakker95-2}) As we discussed in
the previous subsection, other causes for the
discrepancy between theory and numerical simulations may be the
fact that the disorder is not small, or that the parameter $\eta$
is not known.

\section{Conclusions}
\label{sec:Conclusions}

In the vicinity of the band center, the physics of localization 
in a quantum wire with chiral symmetry exhibits differences with
respect to the case of a quantum wire in one of the standard
symmetry classes.
The most prominent differences are observed in the localized regime. 
For wires with chiral symmetry (as is the case with off-diagonal
disorder), at the band center,
the statistics of the conductance depends sensitively on the parity
in the number of transmission channels $N$. For odd $N$, the band center
represents a critical point that is characterized by the absence of
exponential localization. The logarithm of the conductance is 
not self-averaging and the mean conductance or its variance decay 
algebraically with the length $L$ of the wire. 
For even $N$, exponential localization takes place 
with a self-averaging localization length $\xi$ that does not depend on the
presence or absence of time-reversal and spin rotation symmetry.
As the energy is tuned away from the band center, the system crosses over
to the standard universality classes: The parity effect disappears and the 
localization length acquires a dependence on the
presence or absence of time-reversal and spin rotation symmetry.
In the presence of time-reversal symmetry, the localization length
$\xi$ for even $N$ is the same with and without chiral symmetry. 
Our numerical simulations indicate that the crossover is
non-monotonous: in the crossover between the chiral and standard
symmetry classes, $\xi$ differs from the values in the pure
symmetry classes. A complete theoretical description of this crossover
is still lacking.

In the diffusive regime, the differences between the chiral and
standard universality classes are less pronounced. They show up
in quantum interference corrections to the classical (Drude)
conductance, which is the same in both cases.
We have found that in the chiral classes, weak-localization 
corrections to the mean conductance $\langle g \rangle$ and, 
more generally, to the density of transmission eigenvalues vanish
at the band center. This is the quasi-one-dimensional counterpart
of a similar observation made by Gade and Wegner in their study
of two-dimensional disordered systems with chiral
symmetry.\cite{Gade93}
The conductance fluctuations are twice as large
at the band center relative to the standard universality classes,
compatible with the presence of an extra symmetry in the system.
We have calculated these quantum interference corrections as a
function of energy, for the entire crossover from the chiral 
universality class at the band center $\varepsilon=0$ to the standard 
unitary classes for $\varepsilon$ far away from $0$.
The theoretical predictions for this crossover agree qualitatively with 
numerical simulations though there remains sizable deviations 
between theory and numerics of the order of 10-30$\%$.

While the chiral symmetry classes have received an enormous
amount of theoretical attention (see the introduction of this
paper for a brief summary), there are several hurdles to take before a chiral
quantum wire can be realized in practice. Besides the effect
of electron-electron interactions, which is not taken into account
here, the main obstacle is the fact that the chiral symmetry is very 
fragile, since it is easily broken by, e.g., on-site random 
energies, next-nearest-neighbor hopping, or a small shift of
the chemical potential, which will drive the system away from
the chiral symmetric band center. Our calculation of the 
quantum interference corrections in the crossover from the chiral 
symmetry classes to the standard ones can be seen as a first and
necessary step to tackle the latter obstacle.

\acknowledgements

C.~M.~acknowledges support from the Swiss National Science Foundation.
P.~W.~B.~acknowledges  support by the NSF
under grant nos.\ DMR 94-16910, DMR 96-30064, and DMR 97-14725
for the work done at Harvard. 
The work of A.~F.~was supported by a Grant-in-Aid from Japan Society
for the Promotion of Science (No.~11740199). 
The numerical computations were performed at the Yukawa Institute
Computer Facility.

\appendix

\section{Scaling equations}
\label{sec:The appendix}

In this appendix, we present the scaling equations needed to
calculate the crossover for the weak localization corrections of 
the conductance, shot noise, and the universal conductance fluctuations
of the conductance
as done in 
Sec.~\ref{sec:Crossover between the chiral and standard universality classes}. 
The notation was defined in Eq.~(\ref{eq:def for R})
and we use the short-hands
$$
R  \equiv R_{10    },\qquad
R_2\equiv R_{1010  },\qquad
R_3\equiv R_{101010}.
$$
Equations needed up to corrections of order $N^{0}$ are

\bleq 
\ifpreprintsty\else \renewcommand{\thesection}{\Alph{section}} %
\renewcommand{\theequation}{\Alph{section}\arabic{equation}} \fi %

\begin{eqnarray}
{\gamma\ell\over\beta}\partial_L
\left\langle
R
\right\rangle
&&=
N
\left\langle
 N
-2R
\right\rangle
+
\left\langle
R
\right\rangle^2
+{2-\beta\over\beta}
\left\langle
  N
-R^{\ }_{00}
-2R
-R^{* }_{00}
+R^{\ }_{1000}
+R^{* }_{1000}
+R^{\ }_2
\right\rangle,
\\&&\nonumber\\
{\gamma\ell\over\beta}\partial_L
\left\langle
R^{\ }_2
\right\rangle&&=
4\left\langle
 R
-R^{\ }_2
\right\rangle
\left(
N
-\left\langle
R
\right\rangle
\right)
\\
&&
+{2-\beta\over\beta} 
\left\langle
  R^{\ }_{00}
+4R
+ R^{* }_{00}
-8R^{\ }_2
-4R^{\ }_{1000}
-4R^{* }_{1000}
+4R^{\ }_3
+2R^{\ }_{101000}
+2R^{* }_{101000}
+ R^{\ }_{100100}
+ R^{* }_{100100}
\right\rangle,
\nonumber\\&&\nonumber\\
{\gamma\ell\over\beta}\partial_L
{\rm var}\, R
&&=
4
\left(
\left\langle
 R
\right\rangle
-N
\right)
{\rm var}\, R
+\left\langle
 R^{\ }_{00}
+R^{* }_{00}
+2R
-2R^{\ }_{1000}
-2R^{* }_{1000}
-4R^{\ }_2
+R^{\ }_{100100}
+ R^{* }_{100100}
+2R^{\ }_3
\right\rangle
\nonumber\\
&&
+{2-\beta\over\beta}
\left\langle
  R^{\ }_{00}
+ R^{* }_{00}
+2R
-2R^{\ }_{1000}
-2R^{* }_{1000}
-4R^{\ }_2
+ R^{\ }_{100100}
+ R^{* }_{100100}
+2R^{\ }_3
\right\rangle.
\end{eqnarray}
Equations needed up to corrections of order $N$ are
\begin{eqnarray}
{\gamma\ell\over\beta}\partial_L
\left\langle
R^{\ }_{00}
\right\rangle&&=
{4{i}\varepsilon\gamma\ell\over\beta}
\left\langle
R^{\ }_{00}
\right\rangle
+N
\left\langle
N-2R^{\ }_{00}
\right\rangle
+ 
\left\langle
R^{\ }_{00}
\right\rangle^2,
\\&&\nonumber\\
{\gamma\ell\over\beta}\partial_L
\left\langle
R^{\ }_{1000}
\right\rangle&&=
{4{i}\varepsilon\gamma\ell\over\beta} 
\left\langle
R^{\ }_{1000}
\right\rangle
+N
\left\langle
 2R^{\ }_{00}
+2R
-4R^{\ }_{1000}
\right\rangle
-
\left\langle
  R^{\ }_{00}
+2R
-2R^{\ }_{1000}
\right\rangle
\left\langle
R^{\ }_{00}
\right\rangle
-
\left\langle
  R
-2R^{\ }_{1000}
\right\rangle
\left\langle
R
\right\rangle,
\\&&\nonumber\\
{\gamma\ell\over\beta}
\partial_L
\left\langle
R^{\ }_{1100}
\right\rangle&&=
N
\left\langle
 R^{\ }_{00}
+2R
+R^{* }_{00}
-4R^{\ }_{1100}
\right\rangle
- 
\left\langle
 2R
-R^{\ }_{1100}
\right\rangle
\left\langle
R^{\ }_{00}
\right\rangle
-
\left\langle
 2R^{* }_{00}
-2R^{\ }_{1100}
\right\rangle
\left\langle
R
\right\rangle
+ 
\left\langle
R^{\ }_{1100}
\right\rangle
\left\langle
R^{* }_{00}
\right\rangle,
\\&&\nonumber\\
{\gamma\ell\over\beta}\partial_L
\left\langle
R^{\ }_3
\right\rangle&&=
3\left(
2N
\left\langle
 R^{\ }_2
-R^{\ }_3
\right\rangle
+
\left\langle
  R
-4R^{\ }_2
+2R^{\ }_3
\right\rangle
\left\langle
R
\right\rangle
+
\left\langle
R^{\ }_2
\right\rangle^2
\right),
\\&&\nonumber\\
{\gamma\ell\over\beta}\partial_L
\left\langle
R^{\ }_{101000}
\right\rangle&&=
{4{i}\varepsilon\gamma\ell\over\beta}
\left\langle
R^{\ }_{101000}
\right\rangle
+2\left(
N
\left\langle
 2R^{\ }_{1000}
+R^{\ }_2
-3R^{\ }_{101000}
\right\rangle
+ \left\langle
 R
-R^{\ }_{1000}
-R^{\ }_2
+R^{\ }_{101000}
\right\rangle
\left\langle
R^{\ }_{00}
\right\rangle
\right)
\nonumber\\
&&
+
\left\langle
  R
-6R^{\ }_{1000}
-2R^{\ }_2
+4R^{\ }_{101000}
\right\rangle
\left\langle
R
\right\rangle
+
\left\langle
  R^{\ }_{1000}
+2R^{\ }_2
\right\rangle
\left\langle
R^{\ }_{1000}
\right\rangle,
\\&&\nonumber\\
{\gamma\ell\over\beta}\partial_L
\left\langle
R^{\ }_{100100}
\right\rangle&&=
{4{i}\varepsilon\gamma\ell\over\beta}
\left\langle
R^{\ }_{100100}
\right\rangle
+2N
\left\langle
 2R^{\ }_{1000}
+ R^{\ }_{1100}
-3R^{\ }_{100100}
\right\rangle
+ 
\left\langle
  R^{\ }_{00}
-4R^{\ }_{1000}
+2R^{\ }_{100100}
\right\rangle
\left\langle
R^{\ }_{00}
\right\rangle
\nonumber\\
&&
+2
\left\langle
  R
-2R^{\ }_{1000}
-2R^{\ }_{1100}
+2R^{\ }_{100100}
\right\rangle
\left\langle
R
\right\rangle
+2
\left\langle
R^{\ }_{1000}
\right\rangle^2
+
\left\langle
R^{\ }_{1100}
\right\rangle^2.
\end{eqnarray}

\eleq 

\ecols

\end{document}